\documentclass[a4paper,11pt]{article}
\pdfoutput=1 

\usepackage{jheppub} 

\usepackage[T1]{fontenc} 

\usepackage{graphicx} 
\usepackage{dcolumn}  
\usepackage{array,graphicx,grffile,tabularx,float,color,comment,setspace,fancyhdr,bm,amsmath,upgreek,multirow,empheq,indentfirst,tikz,threeparttable}
\usepackage{lineno}
\usepackage{orcidlink}

\begin{document} 

\preprint{\vbox{
		     \hbox{Belle Preprint 2022-09}
                        \hbox{KEK Preprint 2022-8}
}}

\title{\boldmath Measurement of two-photon decay width of $\chi_{c2}(1P)$ in
$ \gamma \gamma \rightarrow \chi_{c2}(1P)  \rightarrow J/\psi\gamma$ at Belle}

\collaborationImg{\includegraphics[width=2cm]{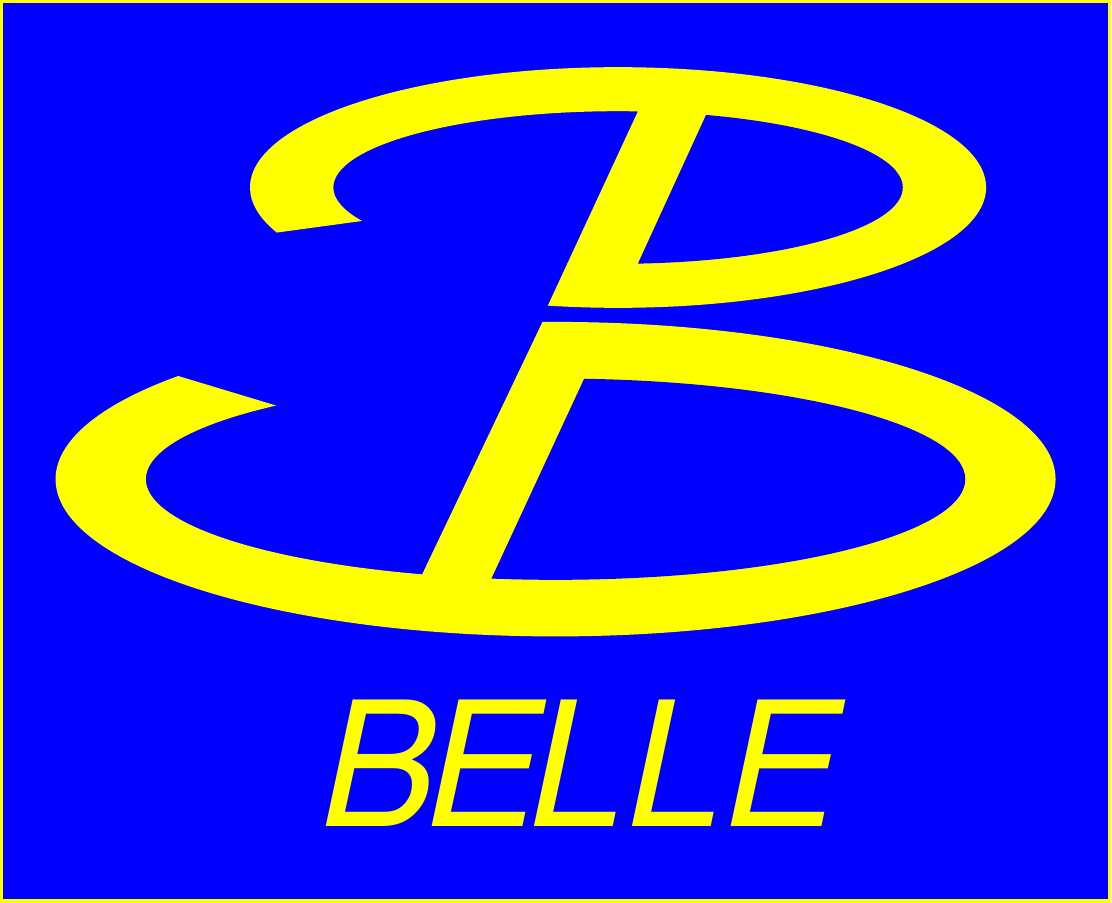}}


\collaboration{The Belle Collaboration}
  \author{Y.~Seino\,\orcidlink{0000-0002-8378-4255},} 
  \author{K.~Hayasaka\,\orcidlink{0000-0002-6347-433X},} 
  \author{S.~Uehara\,\orcidlink{0000-0001-7377-5016},} 
  \author{I.~Adachi\,\orcidlink{0000-0003-2287-0173},} 
  \author{H.~Aihara\,\orcidlink{0000-0002-1907-5964},} 
  \author{S.~Al~Said\,\orcidlink{0000-0002-4895-3869},} 
  \author{D.~M.~Asner\,\orcidlink{0000-0002-1586-5790},} 
  \author{H.~Atmacan\,\orcidlink{0000-0003-2435-501X},} 
  \author{T.~Aushev\,\orcidlink{0000-0002-6347-7055},} 
  \author{R.~Ayad\,\orcidlink{0000-0003-3466-9290},} 
% ...

\abstract{
We report the measurement of the two-photon decay width of $\chi_{c2}(1P)$
in two-photon processes at the Belle experiment.
We analyze the process $ \gamma \gamma \rightarrow \chi_{c2}(1P) \rightarrow J/\psi\gamma$, 
$J/\psi \rightarrow \ell^{+} \ell^{-}$ $(\ell = e \ {\rm or} \ \mu)$
using a data sample of 971~fb$^{-1}$ collected with 
the Belle detector at the KEKB $e^{+} e^{-}$ collider.
In this analysis, 
the product of the two-photon decay width of $\chi_{c2}(1P)$ and the branching fraction is determined to be $\Gamma_{\gamma \gamma}(\chi_{c2}(1P)) \mathcal{B}( \chi_{c2}(1P) \rightarrow J/\psi \, \gamma )\mathcal{B}( J/\psi \rightarrow \ell^+\ell^- ) = {\rm 14.8}$ $\pm$ ${\rm 0.3}({\rm stat.})$ $\pm$ ${\rm 0.7}({\rm syst.})$~eV,
which corresponds to $ \Gamma_{\gamma \gamma}(\chi_{c2}(1P)) $
 = 653  $\pm$ 13(stat.) $\pm$ 31(syst.) $\pm$ 17(B.R.)~eV, where
the third uncertainty is from $\mathcal{B}( \chi_{c2}(1P) \rightarrow J/\psi \, \gamma )$ and $\mathcal{B}( J/\psi \rightarrow \ell^+\ell^- )$.
}

\maketitle
\flushbottom

\section{Introduction}

The two-photon decay widths ($\Gamma_{\gamma\gamma}$) of mesonic states
provide important information 
for testing models based on quantum chromodynamics (QCD), 
which describes quark-antiquark systems.
In particular, it is important to measure the two-photon decay widths of a 
$P$-wave charmonium,
whose description is at the intersection between perturbative and non-perturbative QCD.
Predictions of the two-photon decay width of $\chi_{c2}(1P)$ ($\Gamma_{\gamma\gamma}(\chi_{c2}(1P))$)
have a wide range of values between 280~eV and 930~eV
in various theoretical calculations~\cite{nonrelative_2,relativistic_corrections_1,relativistic_quark_model_2,potential,relativistic_quark_model_1,rigorous_QCD_2,rigorous_QCD_1,nonrelativistic_QCD_factorization_framework,two-body_Dirac_equations_of_constraint_dynamics,effective_Lagrangian,light_front}.
Therefore, precise measurements will help to improve our understanding of quarkonium states. \\
\indent
Several experiments~\cite{CLEOc,bes3,previous_research,CLEO3} have reported the measurement of $\Gamma_{\gamma\gamma}(\chi_{c2}(1P))$.
In general, two approaches have been used to determine 
$\Gamma_{\gamma\gamma}(\chi_{c2}(1P))$,
either measuring the two-photon decay $\chi_{c2}(1P) \rightarrow \gamma \gamma$
or two-photon collisions $\gamma \gamma \rightarrow \chi_{c2}(1P)$.
In the former approach, CLEO-c and \mbox{BES~I\hspace{-.1em}I\hspace{-.1em}I}
data give $\Gamma_{\gamma \gamma}(\chi_{c2}(1P)) = 555 \pm 58 \pm 32 \pm 28 $~eV~\cite{CLEOc}
and $\Gamma_{\gamma \gamma}(\chi_{c2}(1P)) = 586 \pm 16 \pm 13 \pm 29$~eV~\cite{bes3}, respectively,\footnotemark[1]\footnotetext[1]{We recalculate the values in CLEO-c and \mbox{BES~I\hspace{-.1em}I\hspace{-.1em}I} using $\mathcal{B}( \psi(2S) \rightarrow \chi_{c2}(1P) \gamma ) = (9.52 \pm 0.20) \%$ and $\Gamma_{\chi_{c2}(1P)} = 1.97 \pm 0.09$~MeV from PDG \cite{pdg2020}.}
using the production and decay processes $\psi(2S) \rightarrow \chi_{c2}(1P) \gamma$, $\chi_{c2}(1P) \rightarrow \gamma \gamma$.
The uncertainties are statistical, systematic and from $\mathcal{B}(\psi(2S) \rightarrow \chi_{c2}(1P) \gamma)$ and the total width of $\chi_{c2}(1P)$ ($\Gamma_{\chi_{c2}(1P)}$), respectively.
Hereafter, the first and second uncertainties are statistical and systematic.
In the latter approach, Belle and \mbox{CLEO~I\hspace{-.1em}I\hspace{-.1em}I}
data give $\Gamma_{\gamma \gamma}(\chi_{c2}(1P)) = 596 \pm 58 \pm 48 \pm 16$~eV~\cite{previous_research}
and $\Gamma_{\gamma \gamma}(\chi_{c2}(1P)) = 582 \pm 59 \pm 50 \pm 15$~eV~\cite{CLEO3},
respectively,\footnotemark[2]\footnotetext[2]{We recalculate the values in Belle and \mbox{CLEO~I\hspace{-.1em}I\hspace{-.1em}I} using $\mathcal{B}( \chi_{c2}(1P) \rightarrow J/\psi \, \gamma )$ = (19.0 $\pm$ 0.5)\% and
$( J/\psi \rightarrow \ell^+\ell^- )$ = (11.93 $\pm$ 0.05)\% from PDG \cite{pdg2020}.} using $ \gamma \gamma \rightarrow \chi_{c2}(1P) \rightarrow J/\psi\gamma$ in two-photon processes
at $e^{+}e^{-}$ colliders.
The third uncertainty is from $\mathcal{B}( \chi_{c2}(1P) \rightarrow J/\psi \, \gamma )$ and $\mathcal{B}( J/\psi \rightarrow \ell^+\ell^- )$.
At present, the precision of the experimental value of $\Gamma_{\gamma\gamma}(\chi_{c2}(1P))$
using two-photon production is much lower than the value measured in two-photon decay.\\
\indent
In this study, we report an updated measurement of $\Gamma_{\gamma\gamma}(\chi_{c2}(1P))$
in the analysis of $ \gamma \gamma \rightarrow \chi_{c2}(1P) \rightarrow J/\psi\gamma$,
$J/\psi \rightarrow \ell^{+} \ell^{-}$ $(\ell = e \ {\rm or} \ \mu)$
using a data sample of 971~fb$^{-1}$ at Belle.
The previous measurement by Belle~\cite{previous_research}  was performed using a data sample of 32.6~fb$^{-1}$. The increased precision of this measurement will be crucial to
check the consistency between the two approaches of measuring $\Gamma_{\gamma\gamma}(\chi_{c2}(1P))$
and to test the applicability of the existing theoretical models~\cite{nonrelative_2,relativistic_corrections_1,relativistic_quark_model_2,potential,relativistic_quark_model_1,rigorous_QCD_2,rigorous_QCD_1,nonrelativistic_QCD_factorization_framework,two-body_Dirac_equations_of_constraint_dynamics,effective_Lagrangian,light_front}.\\
\indent
The measurement principle of $\Gamma_{\gamma\gamma}(\chi_{c2}(1P))$
in this analysis is as follows.
The production cross section of a resonance $R$ via two-photon processes at $e^{+} e^{-}$ colliders is given by
\begin{equation}
\label{eq:eq1}
\sigma(e^{+}e^{-} \rightarrow e^{+}e^{-}R) = \int \sigma(\gamma \gamma \rightarrow R;W)L_{\gamma \gamma}(W)dW,
\end{equation}
where $W$ is the energy of $R$ in the center-of-mass (c.m.) frame of the two photons emitted from
the $e^{+} e^{-}$ beam,
$\sigma(\gamma \gamma \rightarrow R;W)$ is the production cross section
of $R$ for two-photon collisions at the energy $W$
and $L_{\gamma \gamma}(W)$ is the luminosity function~\cite{epa}, 
which is defined as the probability density of two-photon emission from the $e^{+} e^{-}$ beam,
with the energy $W$.
If the total width of the resonance is sufficiently small compared with its mass,
Eq.~\ref{eq:eq1} can be expressed as~\cite{epa}
\begin{equation}
\label{eq:eq3}
\sigma(e^{+}e^{-} \rightarrow e^{+}e^{-}R) = 4\pi^{2} (2J+1) \frac{L_{\gamma \gamma}(m_{R})\Gamma_{\gamma \gamma}^{R}}{m_{R}^{2}},
\end{equation}
where $J$, $\Gamma_{\gamma \gamma}^{R}$ and $m_{R}$ are the spin quantum numbers,
the two-photon decay width
and the mass of $R$, respectively.
From the observed number of events,
the two-photon decay width of the resonance $R$ can be determined from Eq.~\ref{eq:eq3} as:
\begin{equation}
\label{eq:eq5}
\Gamma_{\gamma \gamma}^{R} \mathcal{B}(R \rightarrow \textrm{final} \ \textrm{state}) = \frac{m_{R}^{2} N_{R}}{4\pi^{2} (2J+1) ({\int \mathcal{L} dt}) \eta L_{\gamma \gamma}(m_{R})},
\end{equation}
where $N_{R}$, ${\int \mathcal{L} dt}$, $\eta$ and $\mathcal{B}(R \rightarrow \textrm{final} \ \textrm{state})$ are
the observed number of $R$ events in the two-photon process,
the integrated luminosity at $e^{+}e^{-}$ collisions,
the detection efficiency and the branching ratio, respectively.

\section{The Belle detector and data sample}

We use a data sample, collected with the Belle detector~\cite{belledetector1,belledetector2}
at the KEKB $e^{+} e^{-}$ asymmetric-energy collider~\cite{KEKB1,KEKB2},
corresponding to a total integrated luminosity of 971~fb$^{-1}$
collected at or near the $\Upsilon (1S)$, $\Upsilon (2S)$, $\Upsilon (3S)$, $\Upsilon (4S)$ and $\Upsilon (5S)$ resonances.
The Belle detector is a large-solid-angle magnetic
spectrometer that consists of a silicon vertex detector (SVD),
a 50-layer central drift chamber (CDC), an array of
aerogel threshold Cherenkov counters (ACC),
a barrel-like arrangement of time-of-flight
scintillation counters (TOF), and an electromagnetic calorimeter
comprised of CsI(Tl) crystals (ECL) located inside 
a super-conducting solenoid coil that provides a 1.5~T
magnetic field.  An iron flux-return located outside of
the coil is instrumented to detect $K_L^0$ mesons and to identify
muons (KLM).
Events are selected with an OR of 
the two-track trigger and the ``HiE'' trigger,
providing a 98.3\% trigger efficiency.
The condition for the two-track trigger is CDC hits from two charged tracks
with either associated ECL clusters with an energy sum $> 0.5$~GeV,
or associated KLM hits.
The condition for the HiE trigger
is an energy sum $> 1.0$~GeV over all ECL clusters.\\
\indent
We use the TREPS generator~\cite{treps} for Monte Carlo (MC) simulation
of the two-photon process.
For the signal, we generate the processes
$e^{+}e^{- }  \rightarrow e^{+}e^{- }\chi_{c2}(1P)$, $\chi_{c2}(1P) \rightarrow J/\psi\gamma$, $J/\psi \rightarrow \ell^{+} \ell^{-}$ $(\ell = e \: {\rm or} \: \mu)$.
The effect of $J/\psi$ radiative decays to $\ell^{+} \ell^{-} \gamma$ is simulated using PHOTOS~\cite{photos}.
The generated events are processed by the full detector simulation based on GEANT3~\cite{geant3}.
TREPS is also used to calculate the luminosity function based on the equivalent photon approximation~\cite{epa}.
When we estimate the overall signal detection efficiency for the determination of
$\Gamma_{\gamma\gamma}(\chi_{c2}(1P))$, 
the signal MC is generated under the following condition.
We prepare the signal MC at the $e^{+}e^{-}$ c.m.\ energy corresponding to the $\Upsilon (4S)$ resonance.
This c.m.\ energy comprises the majority of the used data sample and differences of the beam energy within the data sample are estimated to have an effect of less than 0.1\% on the cross-section and the detection efficiency
of the signal processes.
Similarly, we use a mass of 3.556 GeV/$c^{2}$ for $\chi_{c2}(1P)$ and
we neglect the finite width, since its effect on the cross-section and the detection efficiency
is also estimated to be small, about 0.4\%, and these effects are added to the systematics.
For the angular distribution of the decay in the signal MC of $\chi_{c2}(1P)$,
we assume $a_{2}=-0.11$, $a_{3}=0.00$ and a pure $\uplambda=2$ state,
where $a_{2}$, $a_{3}$ and $\uplambda$ are defined as the fractional multipole amplitudes of M2 and E3 transitions
and the helicity of the $\chi_{c2}(1P)$ with respect to the $\gamma$$\gamma$ axis, respectively. (See Appendix~\ref{appendixa} for details.)

\section{Event Selection}

\begin{figure}[htbp]
 \begin{minipage}{1.0\hsize}
  \begin{center}
  \includegraphics[width=1.0\textwidth]{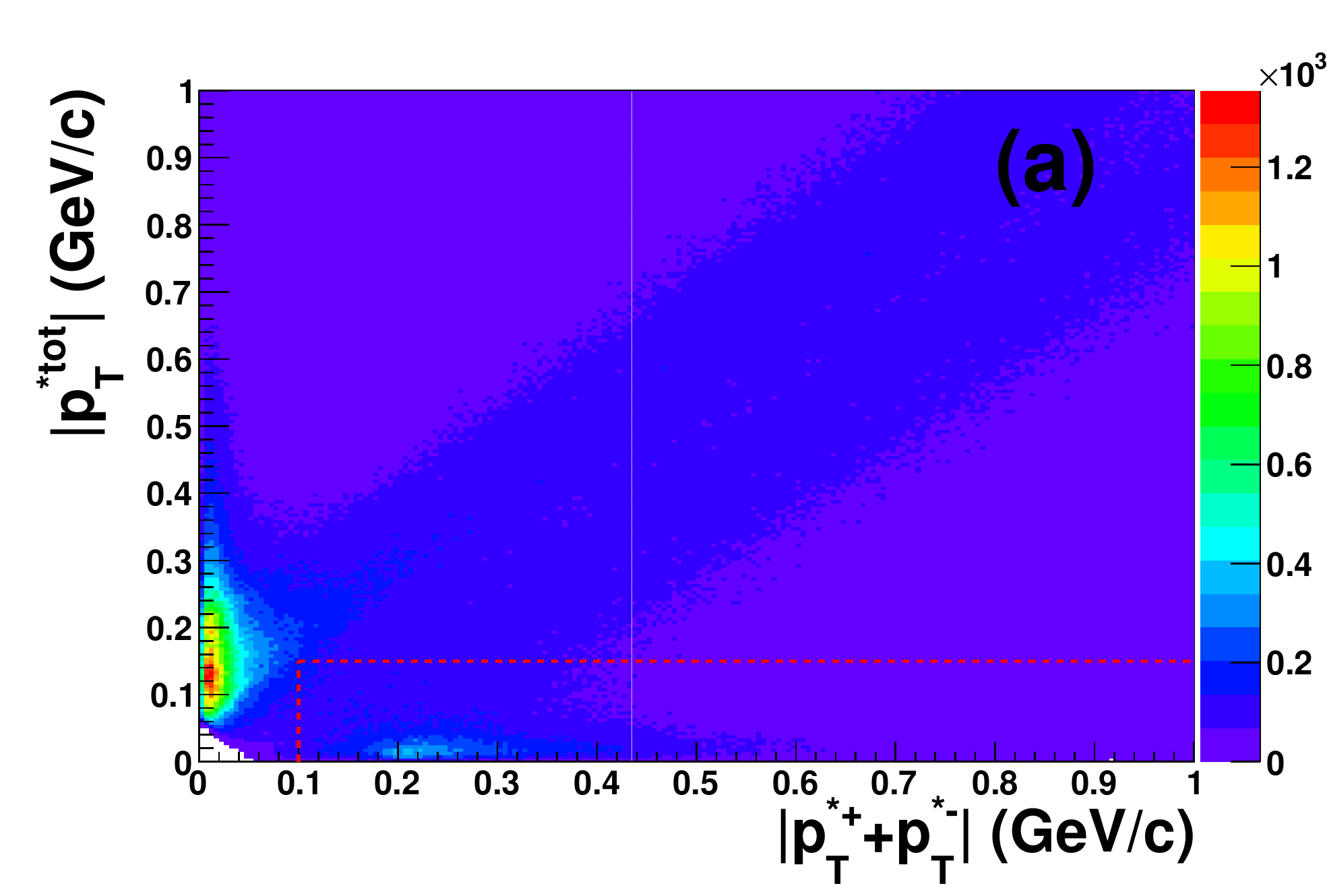}
  \end{center}
 \end{minipage}
 \begin{minipage}{1.0\hsize}
  \begin{center}
  \includegraphics[width=1.0\textwidth]{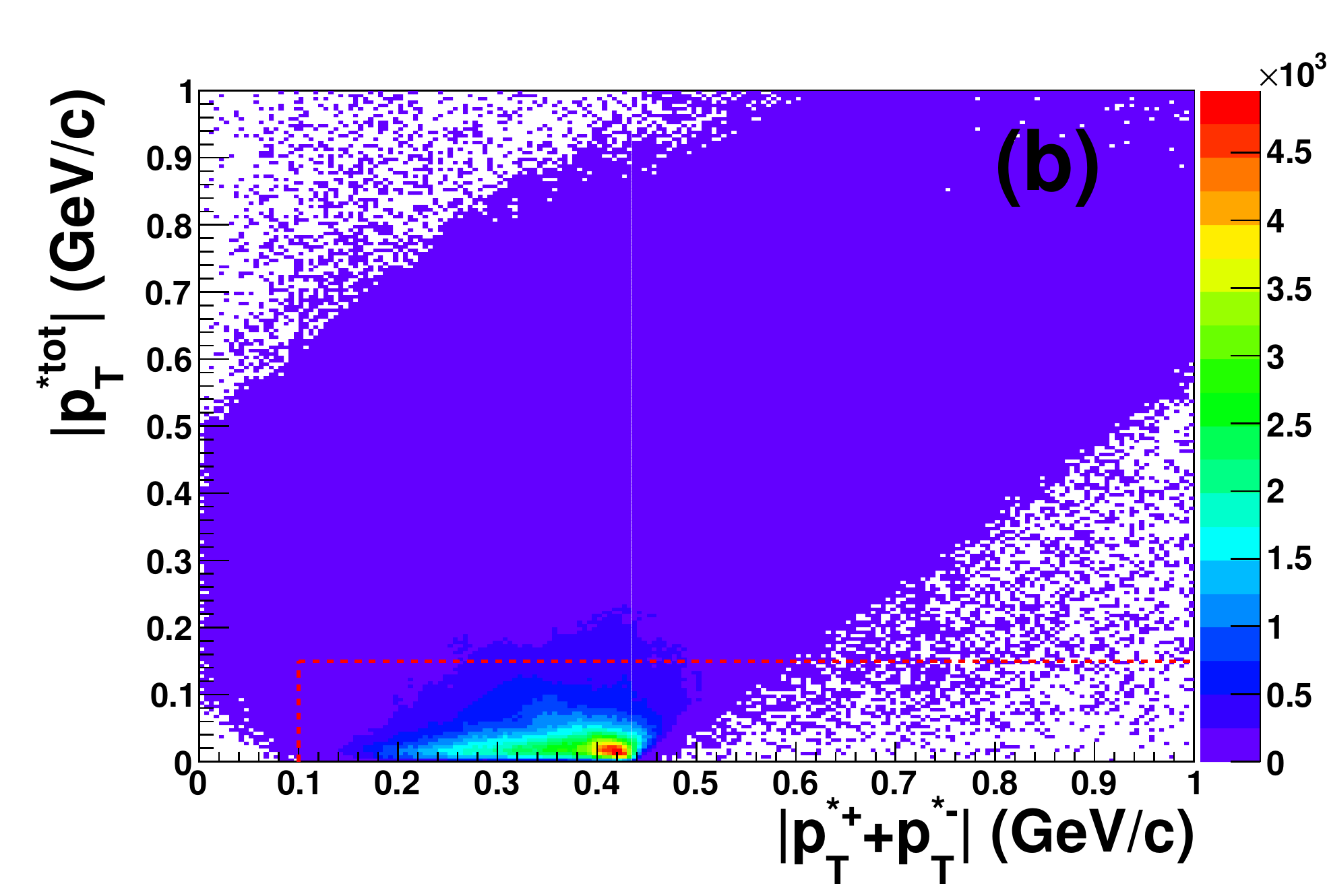}
  \end{center}
 \end{minipage}
  \caption{Scatter plots of $| {\bm p}_{\mathrm T}^{*{\rm tot}} |$ versus $|{\bm p}_{\mathrm T}^{*+} + {\bm p}_{\mathrm T}^{*-}|$ after the event selection except
for the selections using $|{\bm p}_{\mathrm T}^{*+} + {\bm p}_{\mathrm T}^{*-}|$ and $| {\bm p}_{\mathrm T}^{*{\rm tot}} |$: (a) for the data sample and (b) for the signal MC. Dotted red lines show the selected region according to the selections using $|{\bm p}_{\mathrm T}^{*+} + {\bm p}_{\mathrm T}^{*-}|$ and $| {\bm p}_{\mathrm T}^{*{\rm tot}} |$.}
  \label{pt_plot}
 \end{figure}

We select an $e^{+}e^{-}$ or $\mu^{+}\mu^{-}$ pair with
a signal photon to reconstruct $\chi_{c2}(1P)$.
To determine $\Gamma_{\gamma\gamma}(\chi_{c2}(1P))$,
we select quasi-real two photon collisions,
requiring that both recoiling beam particles are undetected (``zero-tag mode'').
This event selection is similar to the previous Belle analysis~\cite{previous_research}.\\
\indent
For the $e^{+}e^{-}$ or $\mu^{+}\mu^{-}$ selection,
we require that only two oppositely charged tracks are present in the event. 
These tracks have to fulfill the following conditions in the laboratory frame:
$-0.47 \leq \cos\theta \leq 0.82$, where $\theta$ is
the polar angle; $p_{\mathrm T} \geq 0.4 $~GeV/$c$, where $p_{\mathrm T}$ is
the transverse momentum; $|dz| \leq$ 3~cm and $dr \leq$ 1~cm,
where $dz$ and $dr$ are the impact parameters relative to the beam interaction 
point (IP) along the $z$ axis
defined as the direction opposite that of the $e^{+}$ beam
and the transverse plane, respectively;
$|\Delta dz| \leq$ 1~cm, where $\Delta dz$ is the difference between 
the $dz$'s of the two tracks;
$\cos\alpha > -0.997$, where $\alpha$ is the opening angle of the two tracks,
to reject the cosmic-ray backgrounds. 
Events are identified as $e^{+}e^{-}$ ($\mu^{+}\mu^{-}$)
if both tracks have $E/pc \geq$ 0.8 ($E/pc \leq$ 0.4),
where $E$ and $p$ are the energy deposit on ECL
and the measured momentum, respectively.
Events rejected by this criterion are mainly charged hadron backgrounds.
In the case the tracks are identified as an $e^{+}e^{-}$ pair,
their momentum is corrected for the effect of bremsstrahlung as explained in the following.
If there are one or more photons which have a total energy between
0.02~GeV and 0.2~GeV within a cone with an angle of 3$^{\circ}$ around
the direction of electron momentum, the energy of these photons is
added to that of the electron track.
This procedure also reduces the width of the $J/\psi$ radiative tail.\\
\indent
For the signal photon from $\chi_{c2}(1P)$ decay,
we require just one cluster in the ECL with an 
energy $E_{\gamma} \geq$ 0.2~GeV under the 
condition that the cluster is isolated from the nearest charged track by an 
angle greater than 18.2$^{\circ}$.\\
\indent
To reconstruct the $\chi_{c2}(1P)$ produced in
the two-photon process in the zero-tag mode,
the following conditions are required.
The scalar sum of the momenta of the two tracks
must be less than 6.0~GeV/$c$ and
the total energy deposited in the ECL must be less than 6.0~GeV.
These requirements reject $e^{+} e^{-}$ annihilation processes.
To reject the initial-state-radiation (ISR) process,
only events with $M^{2}_{\rm rec} >$ 5.0~GeV$^{2}$/$c^{4}$ are selected,
where $M^{2}_{\rm rec}$ is the square of the recoil mass, which is defined as
$M^{2}_{\rm rec} = (E_{\rm beam}^{*}-E_{+-}^{*})^{2}/c^{4} - |{\bm p}_{+-}^{*}|^{2}/c^{2} $.
We define $E_{\rm beam}^{*}$, $E_{+-}^{*}$ and 
$|{\bm p}_{+-}^{*}|$ as the sum of $e^{+}e^{-}$ beam energies, the sum of the energy of two tracks and
the absolute value of the sum of the momentum vector of two tracks in the c.m.\ frame of the $e^{+}e^{-}$ beam, respectively.
The signal $\chi_{c2}(1P)$ produced in quasi-real two-photon collisions
are selected with a ${\bm p}_{\mathrm T}^{*}$-balance requirement.
This requirement is that the absolute value of the total transverse momentum vector
in the c.m.\ frame of the $e^{+}e^{-}$ beam,
which is defined as $| {\bm p}_{\mathrm T}^{*{\rm tot}}| = |{\bm p}_{\mathrm T}^{*+} + {\bm p}_{\mathrm T}^{*-} + {\bm p}_{\mathrm T}^{*\gamma}|$, must be less than 0.15~GeV/$c$,
where ${\bm p}_{\mathrm T}^{*+}$, ${\bm p}_{\mathrm T}^{*-}$ and ${\bm p}_{\mathrm T}^{*\gamma}$ are the 
transverse momentum vectors of 
$\ell^{+}$, $\ell^{-}$ and
the signal photon, respectively.
Furthermore, we require that $|{\bm p}_{\mathrm T}^{*+} + {\bm p}_{\mathrm T}^{*-}|$
be larger than 0.1~GeV/$c$
to reject $\ell^{+} \ell^{-}$ pair produced in two-photon process with a fake photon.
After the event selection except
for the selection using $|{\bm p}_{\mathrm T}^{*+} + {\bm p}_{\mathrm T}^{*-}|$ and $| {\bm p}_{\mathrm T}^{*{\rm tot}} |$,
the scatter plots of $| {\bm p}_{\mathrm T}^{*{\rm tot}} |$ versus $|{\bm p}_{\mathrm T}^{*+} + {\bm p}_{\mathrm T}^{*-}|$ for the data sample 
and the signal MC are shown in Fig.~\ref{pt_plot}~(a) and (b), respectively.
There are two clusters of events at 
$|{\bm p}_{\mathrm T}^{*+} + {\bm p}_{\mathrm T}^{*-}| \approx 0$~GeV/$c$ and $ | {\bm p}_{\mathrm T}^{*{\rm tot}} | \approx 0$~GeV/$c$
in Fig.~\ref{pt_plot}~(a).
The former corresponds to the $\ell^{+} \ell^{-}$ pair production in two-photon process
with a fake photon.
The latter corresponds to the exclusive $\ell^{+} \ell^{-} \gamma$ final state in two-photon process,
which is the dominant background component in this analysis;
in Fig.~\ref{pt_plot}~(a), the signal events are not clear due to
the large background at this stage of the analysis.
A clear cluster at $ | {\bm p}_{\mathrm T}^{*{\rm tot}} | \approx 0$~GeV/$c$
can be seen in the signal MC events in Fig.~\ref{pt_plot}~(b).
The region corresponding to the selections is drawn by the dotted red lines in Fig.~\ref{pt_plot}~(a) and (b).

\newpage

\section{\boldmath $\chi_{c2}(1P)$ signal extraction}

\begin{figure}[htbp]
   \begin{center}
    \includegraphics[width=1.0\textwidth]{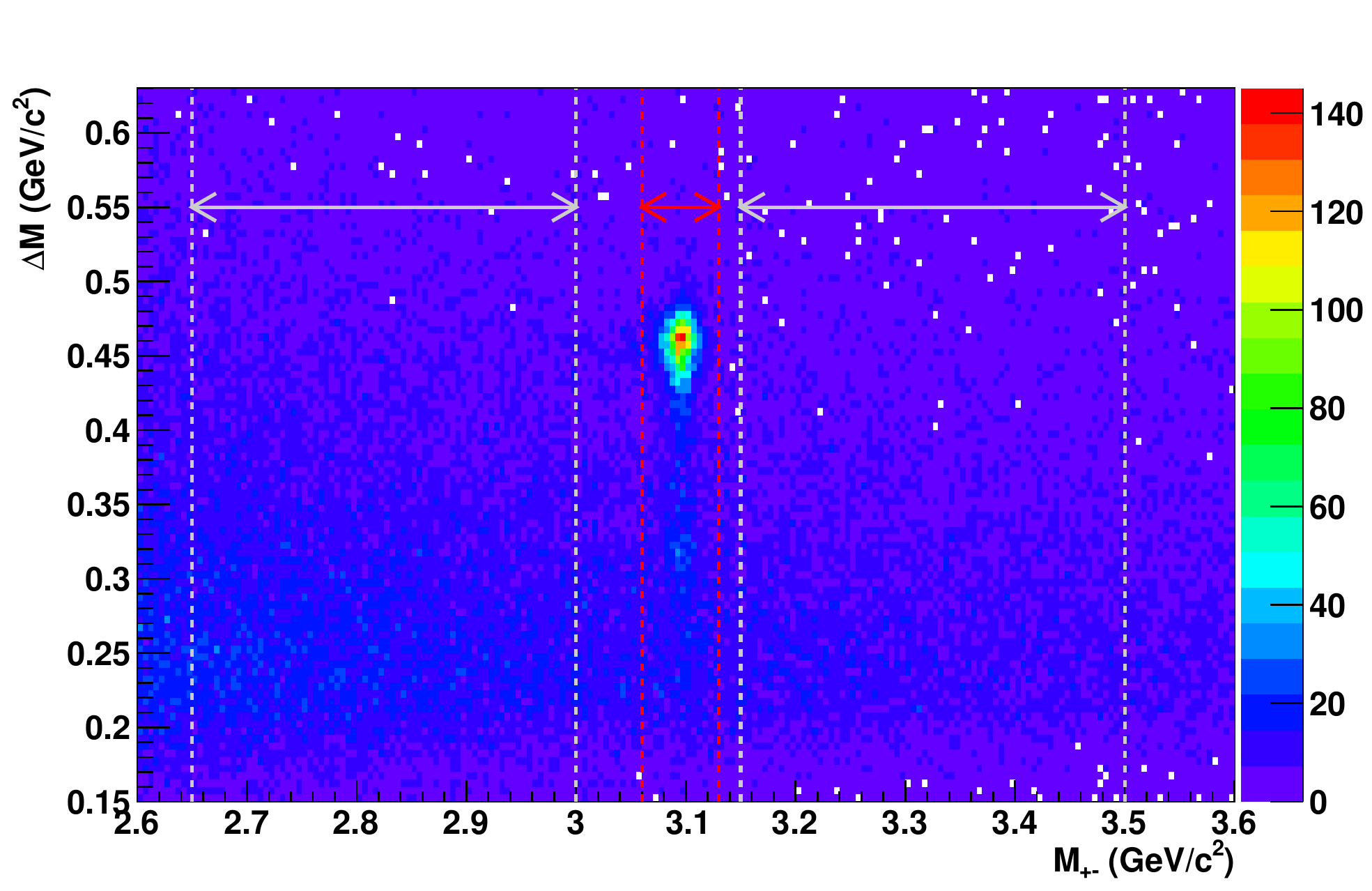}
    \caption{Scatter plot of the invariant mass difference ($\Delta M $) versus the invariant mass of the two tracks ($M_{+-}$) for the data sample after the event selection combined with $e^{+}e^{-}$ and $\mu^{+}\mu^{-}$ pairs. The $J/\psi$ signal region and $J/\psi$ sideband regions are drawn by 
the red and gray dotted line, respectively.}
    \label{inv_mass}
   \end{center}
  \end{figure}

Figure~\ref{inv_mass} shows the scatter plot of 
the invariant mass difference, $\Delta M = M_{+-\gamma} - M_{+-}$, 
versus the invariant mass of the two tracks ($M_{+-}$)
for the data sample after the event selection, where
$M_{+-\gamma}$ is the invariant mass of the selected two tracks and the signal photon candidate.
As expected, a well separated cluster, corresponding to the signal of $\chi_{c2}(1P)  \rightarrow J/\psi \gamma$ is evident in Fig.~\ref{inv_mass}.
To define the $J/\psi$ sideband events and the $J/\psi$ signal candidates,
we set the $J/\psi$ sideband regions
(2.65~GeV/$c^{2}$ $<$ $M_{+-} <$~3.00 GeV/$c^{2}$ and 3.15~GeV/$c^{2}$ $<$ $M_{+-} <$ 3.50~GeV/$c^{2}$)
and the $J/\psi$ signal region (3.06~GeV/$c^{2}$ $\leq$ $M_{+-} \leq$ 3.13~GeV/$c^{2}$), respectively.
To take the radiative tail of the $J/\psi$ signal into account,
the $J/\psi$ sideband regions are defined to be asymmetric with respect to the $J/\psi$ signal region.

\begin{figure}[htbp]
 \begin{minipage}{1.0\hsize}
  \begin{center}
  \includegraphics[width= 0.8\textwidth]{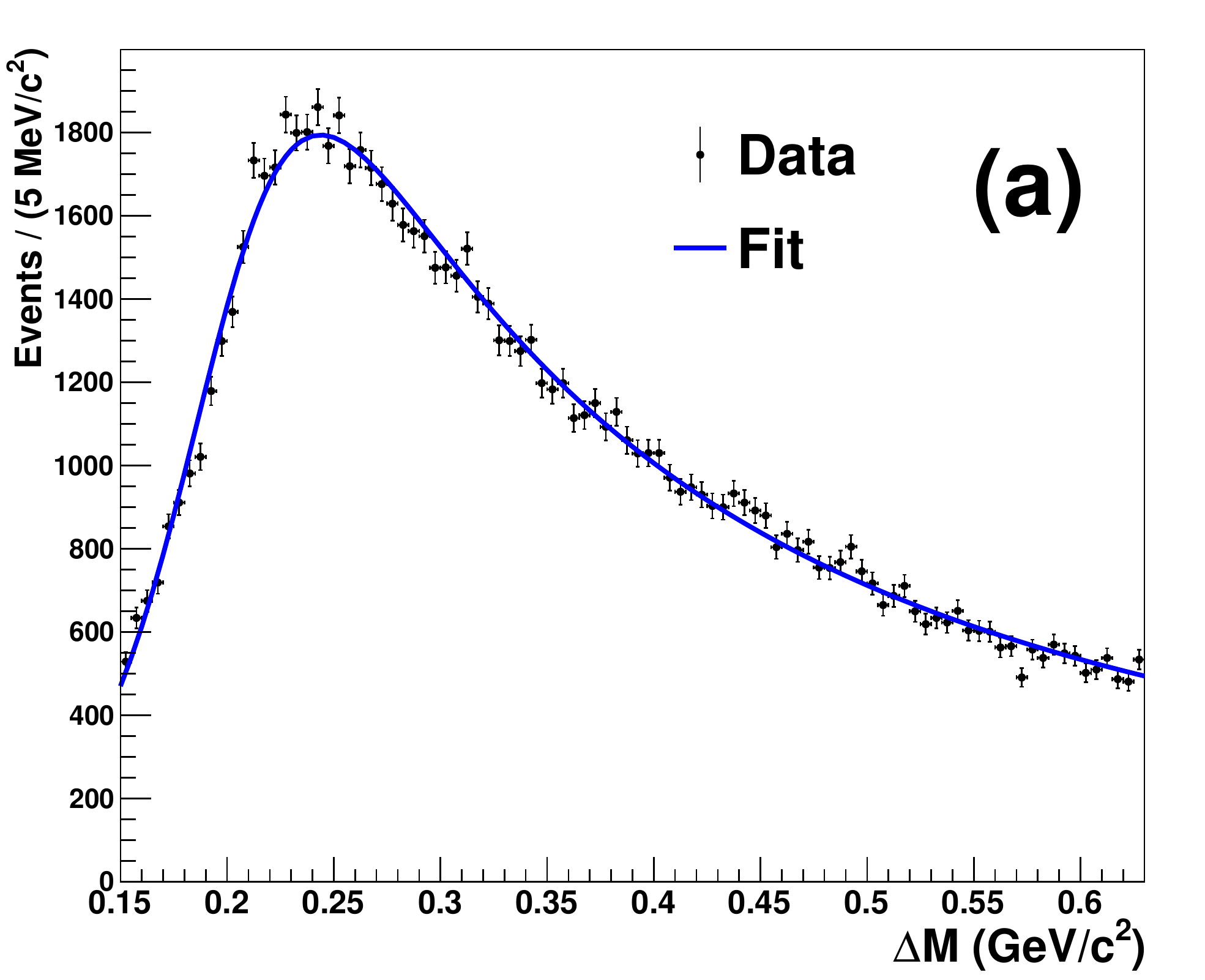}
  \end{center}
\end{minipage}
 \begin{minipage}{1.0\hsize}
  \begin{center}
  \includegraphics[width=0.8\textwidth]{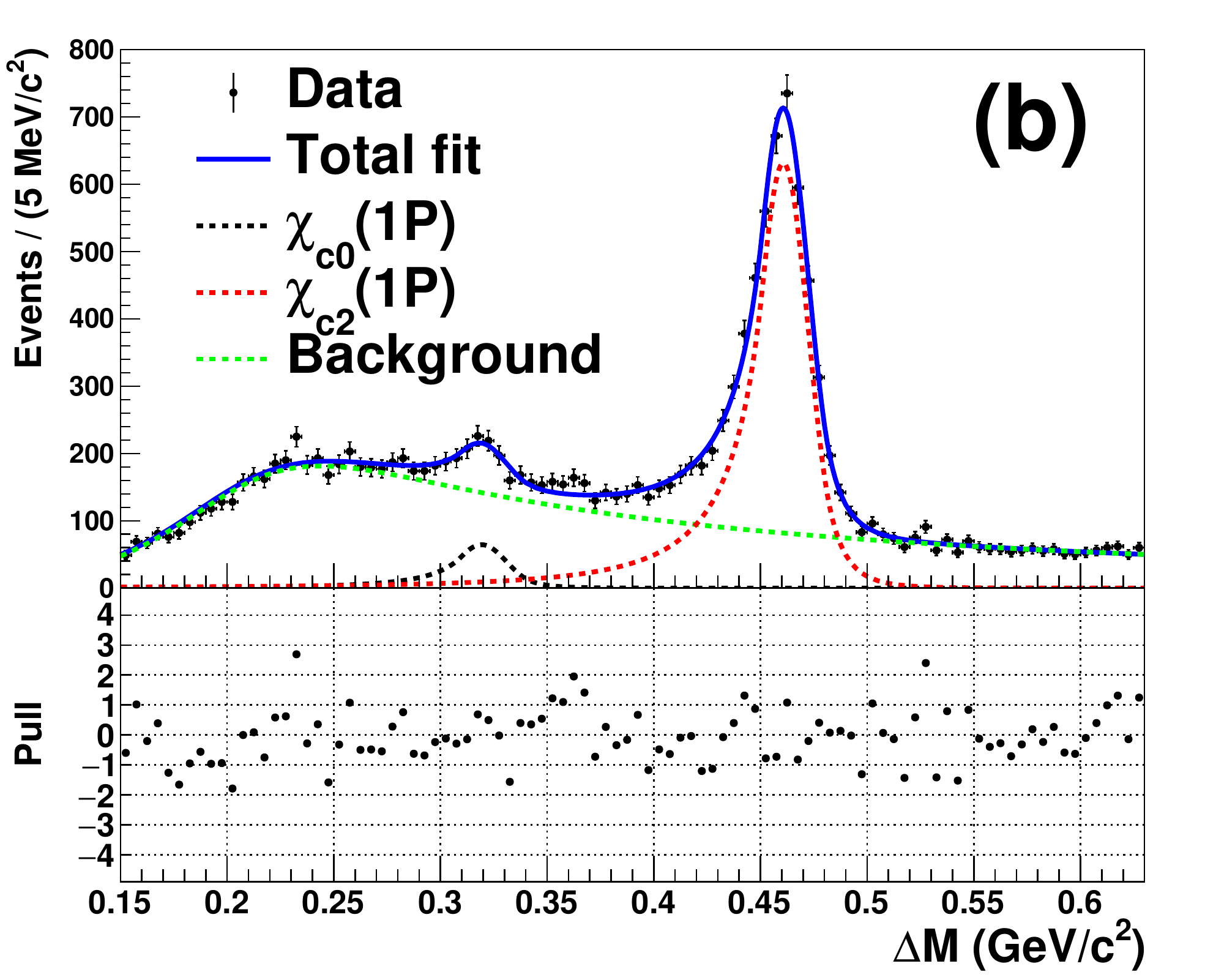}
  \end{center}
 \end{minipage}
  \caption{Mass difference distribution for (a) $J/\psi$ sideband events and (b) $J/\psi$ signal candidates combined with $e^{+}e^{-}$ and $\mu^{+}\mu^{-}$ pairs. The solid blue curves and the black points with error bars show the overall fit result and the data, respectively. The dotted black, red and green curves show the $\chi_{c0}(1P)$, $\chi_{c2}(1P)$ and background components in (b), respectively. The pull plot in (b) shows the residual between 
the overall fit result and the data divided by the uncertainty in each bin.}
  \label{delta_mass}
 \end{figure}

\begin{figure}[htbp]
\begin{tabular}{cc}
 \begin{minipage}{0.5\hsize}
  \begin{center}
  \includegraphics[width= 1.0\textwidth]{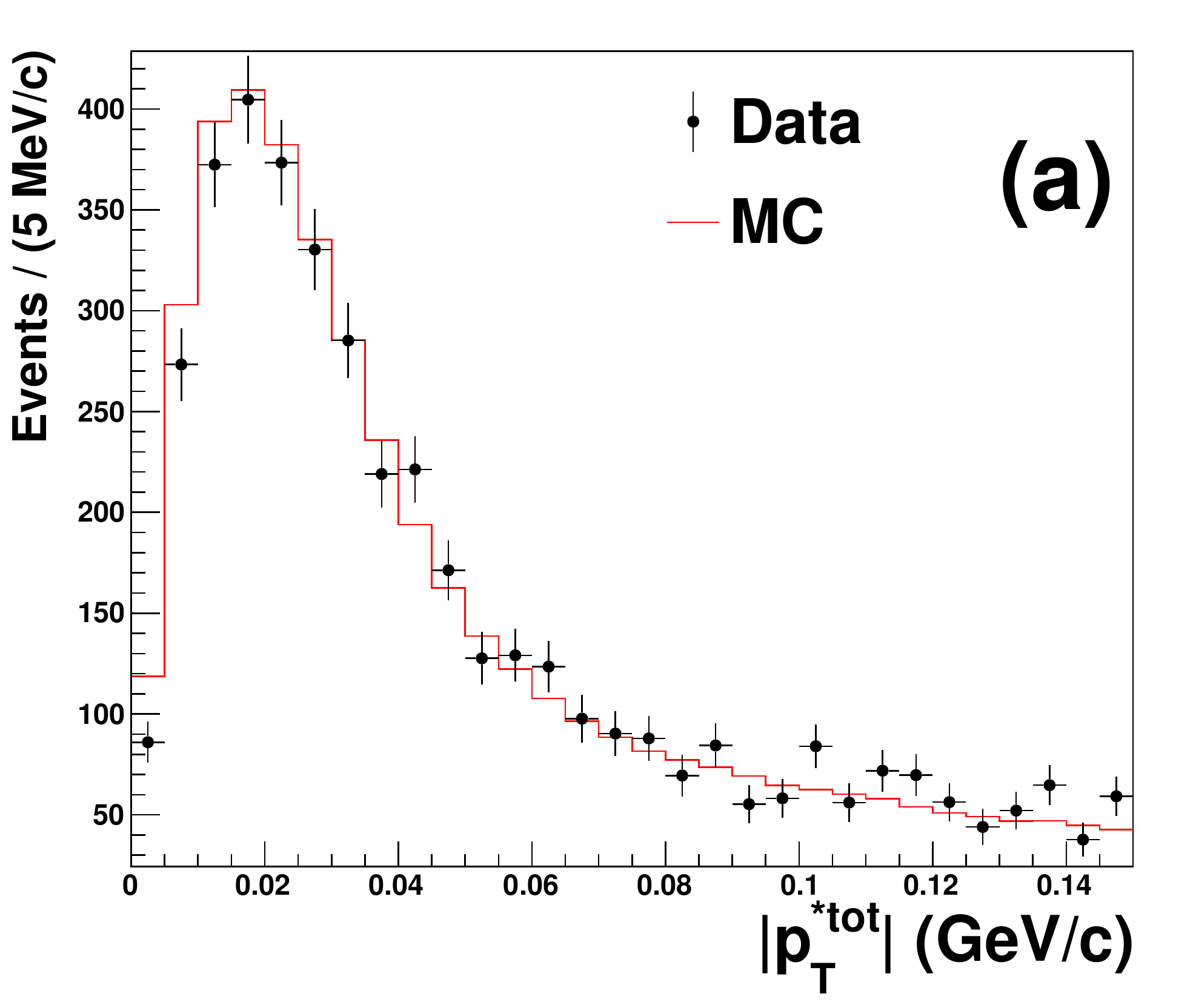}
  \end{center}
\end{minipage}&
 \begin{minipage}{0.5\hsize}
  \begin{center}
  \includegraphics[width=1.0\textwidth]{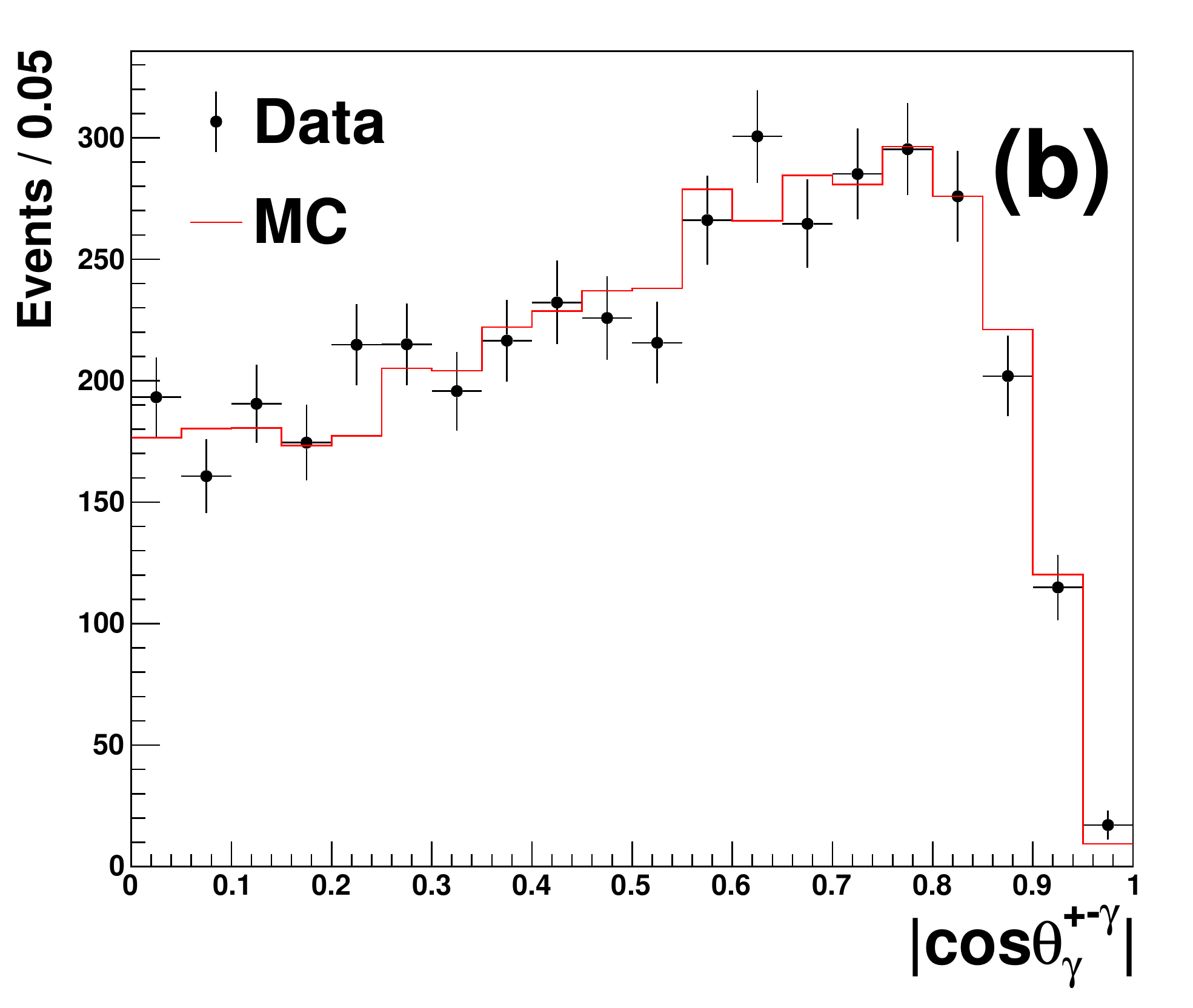}
  \end{center}
 \end{minipage}\\
\begin{minipage}{0.5\hsize}
  \begin{center}
  \includegraphics[width= 1.0\textwidth]{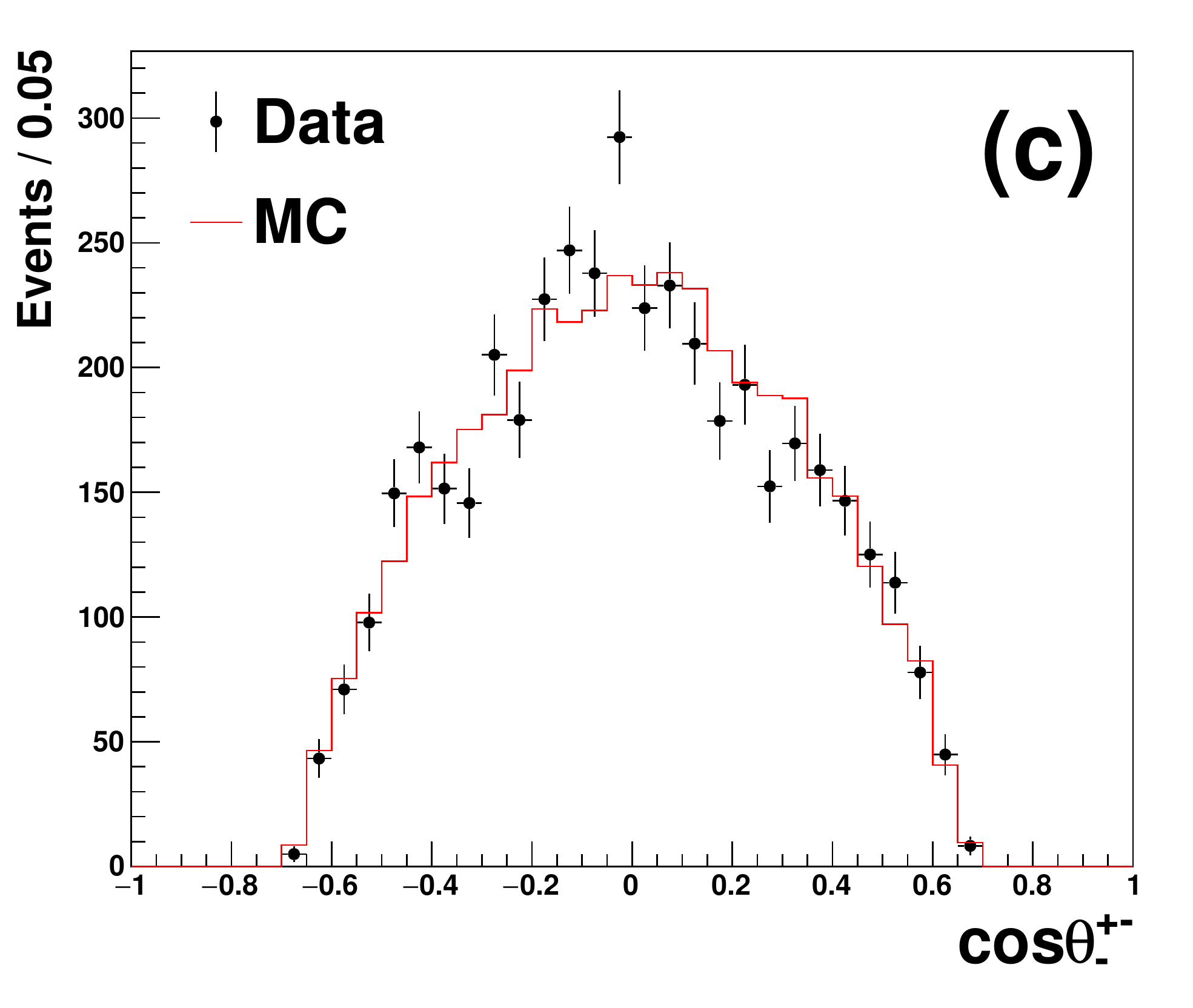}
  \end{center}
\end{minipage}&
 \begin{minipage}{0.5\hsize}
  \begin{center}
  \includegraphics[width=1.0\textwidth]{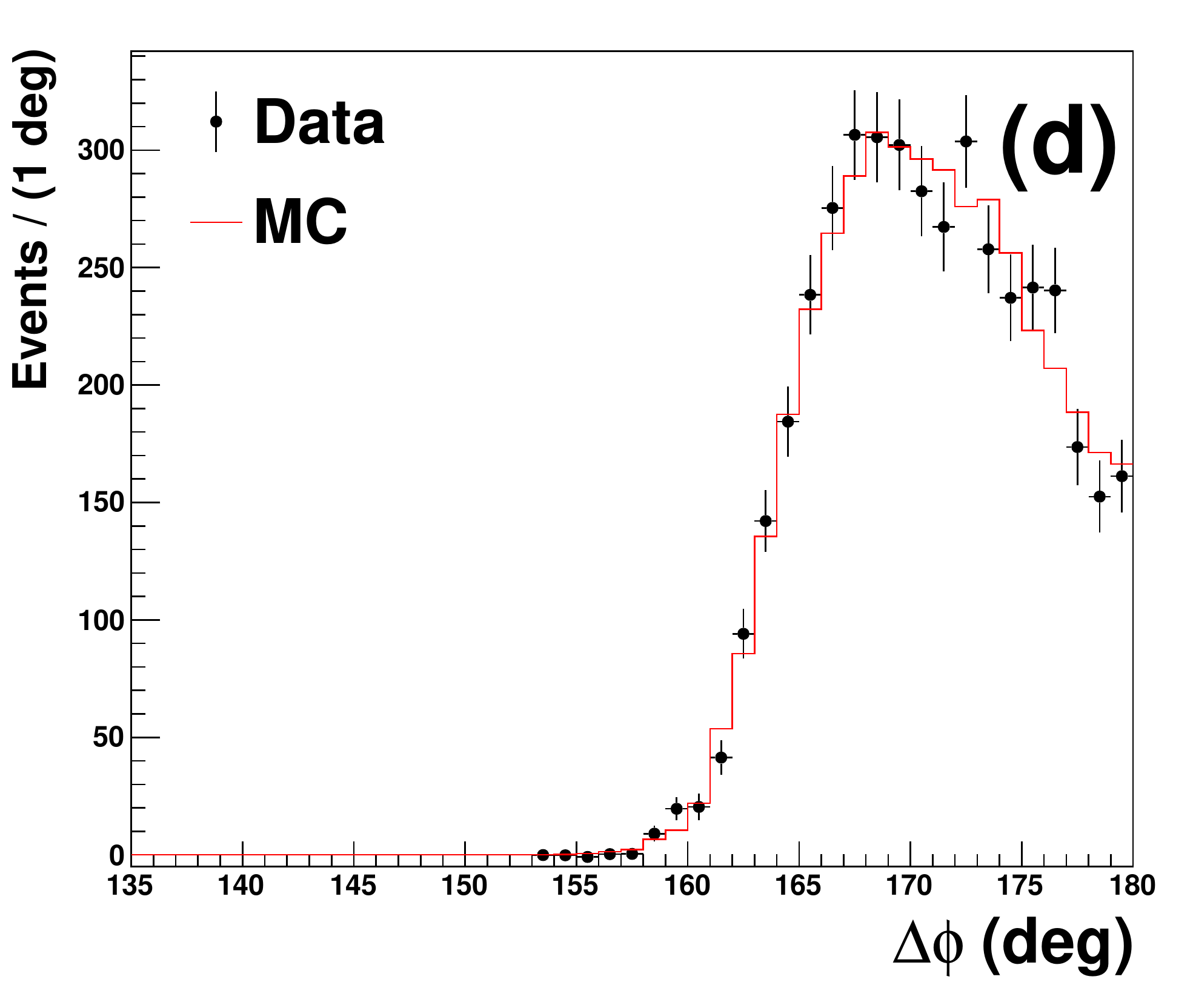}
  \end{center}
 \end{minipage}
\end{tabular}
  \caption{Comparisons of the signal MC (red histogram) with the background sample subtracted data sample estimated from $J/\psi$ sideband events combined with $e^{+}e^{-}$ and $\mu^{+}\mu^{-}$ pairs (black points with error bars) between $\Delta M $ = 0.42~ GeV/$c^{2}$ and 0.49~GeV/$c^{2}$ for (a) $| {\bm p}_{\mathrm T}^{*{\rm tot}} |$; (b) $|\cos\theta^{+-\gamma}_{\gamma}|$; (c) $\cos\theta^{+-}_{-}$; (d) $\Delta \phi$. There are no entries in the $\Delta \phi < 135^{\circ}$.
The signal MC is normalized to the observed number of events in data sample.}
  \label{compare}
 \end{figure}

Non-$J/\psi$ background is expected to be the dominant background component in this analysis,
because spin-1 meson production is suppressed in quasi-real two-photon collisions.
We use the $J/\psi$ sideband events for the data sample to study
the background component.
Figure~\ref{delta_mass}~(a) shows
the mass difference distribution in $J/\psi$ sideband events for the data sample fitted by
the following empirical function from the previous Belle study \cite{previous_research}
using a binned maximum-likelihood fit:
 $A(\Delta M -a)^{-b}/(1 + e^{-c(\Delta M -d)})$, where $a$, $b$, $c$ and $d$ are shape parameters,
and $A$ is the normalization parameter. 
From this fit
we obtain the expected shape of the background component in the $\Delta M $ distribution.
Furthermore,
the number of background events expected in the signal region
is found to be 9966 $\pm$ 32 events by scaling the number of events in the sideband.\\
\indent
In the previous Belle measurement~\cite{previous_research}, 
the number of $\chi_{c2}(1P)$ signal events was estimated by subtracting the number of background events 
from the total in a narrow $\chi_{c2}(1P)$ signal region.
However, this method suffered from a low detection efficiency and a large systematic uncertainty.
In this analysis, we use a fit method to improve these points.
To extract the $\chi_{c2}(1P)$ signal from the $\Delta M $ distribution in $J/\psi$ signal candidates,
we perform a binned extended maximum-likelihood fit
with probability density functions (PDFs) corresponding to $\chi_{c0}(1P)$ and $\chi_{c2}(1P)$
signal and background components.
The signal of $\chi_{c0}(1P)$ is expected to be non-negligible.
On the other hand, the influence of $\chi_{c1}(1P)$ is negligibly small
because $\chi_{c1}(1P)$ is a spin-1 meson suppressed in quasi-real two photon collision.
The ratio of the production and decay process for
$\chi_{c1}(1P)  \rightarrow J/\psi\gamma $ to $\chi_{c2}(1P)  \rightarrow J/\psi\gamma $
is estimated to be only $4 \times 10^{-4}$ in this analysis~\cite{nonrelativistic_QCD_factorization_framework,teramoto}.
A double-sided Crystal Ball function\footnotemark[3]\footnotetext[3]{
$f(x;n_{l},\alpha_{l},n_{r},\alpha_{r},\mu,\sigma)$\\
$= N (n_{l}/|\alpha_{l}|)^{n_{l}} {\rm exp} ( - |\alpha_{l}|^{2}/2)[ (n_{l}/|\alpha_{l}|) - |\alpha_{l}| - (x-\mu)/\sigma)]^{-n_{l}},$ ${\rm if} \ \ (x-\mu)/\sigma \leq - \alpha_{l}$\\
$= N (n_{r}/|\alpha_{r}|)^{n_{r}} {\rm exp} ( - |\alpha_{r}|^{2}/2)[ (n_{r}/|\alpha_{r}|) - |\alpha_{r}| + (x-\mu)/\sigma)]^{-n_{r}},$ ${\rm if} \ \ (x-\mu)/\sigma \geq \alpha_{r}$\\
$=N{\rm exp} [( - (x-\mu)^{2})/(2\sigma^{2})],$ otherwise.\\
$N$ is a normalization parameter. 
$n_{l}$ and $\alpha_{l}$ are parameters of the left-hand side tail,
$n_{r}$ and $\alpha_{r}$ are parameters of the right-hand side tail, and
$\mu$ and $\sigma$ are the peak position and width of the gaussian term.}
is empirically suitable for the $\chi_{c0}(1P)$ and $\chi_{c2}(1P)$ signal shape~\cite{chi_b}.
The tail parameters ($n_{l}$, $\alpha_{l}$, $n_{r}$, $\alpha_{r}$) of the $\chi_{c2}(1P)$ signal PDF
are fixed to the values obtained from the MC study
of the signal, where the total width was taken into account.
The parameters $\mu$ and $\sigma$ in the $\chi_{c2}(1P)$ signal PDF are floated.
The tail parameters and $\sigma$ of the $\chi_{c0}(1P)$ signal PDF are fixed to
those of the $\chi_{c2}(1P)$ signal PDF
since there is only a small contribution from the $\chi_{c0}(1P)$
and the width of the $\chi_{c0}(1P)$ signal shape
is expected to be close to that of the $\chi_{c2}(1P)$ signal shape obtained from the MC study.
Furthermore, the mean of the $\chi_{c0}(1P)$ signal PDF
is constrained to that of the $\chi_{c2}(1P)$ signal PDF
with the mass difference between $\chi_{c0}(1P)$
and $\chi_{c2}(1P)$ taken from the PDG~\cite{pdg2020}.
The background PDF is taken to be the same function as described in the study of $J/\psi$
sideband events with the shape parameters ($a$, $b$, $c$, $d$)
fixed to the values from the fit result shown in Fig.~\ref{delta_mass}~(a).
From a fitter test using toy MC simulations based on 
the shape of signal MCs and the background shape
estimated from the study of $J/\psi$ sideband events,
we confirm that the fit method is stable and has no bias.
Figure~\ref{delta_mass} (b) shows
the mass difference distribution in $J/\psi$ signal candidates for the data sample fitted by 
this method.
The fitted $\chi_{c2}(1P)$ signal and background yields are 5131.2 $\pm$ 97.4
and 10079 $\pm$ 140 events, respectively.
The number of background events estimated from this fit is
consistent with that estimated from the study of the $J/\psi$ sideband events.\\
\indent
Figure~\ref{compare} shows the comparison of the signal MC
with background-subtracted data for events with $\Delta M$ from 0.42~GeV/$c^{2}$ to 0.49~GeV/$c^{2}$
for $| {\bm p}_{\mathrm T}^{*{\rm tot}} |$, $|\cos\theta^{+-\gamma}_{\gamma}|$, $\cos\theta^{+-}_{-}$ and $\Delta \phi$, respectively; $J/\psi$ sidebands are used to estimate the background.
The definition of variables in Fig~\ref{compare} is as follows.
$|\cos\theta^{+-\gamma}_{\gamma}|$ is defined as 
the polar angle of the signal photon in the $\ell^{+} \ell^{-} \gamma$ c.m.\ frame;
$\cos\theta^{+-}_{-}$ is defined as the polar angle of the negatively charged lepton in the $\ell^{+} \ell^{-}$ c.m.\ frame; $\Delta \phi$ is defined as the difference in
the azimuthal angle between the momentum vectors of the two leptons
in the laboratory frame.
The signal MC is normalized to the observed number of events
in the data sample in Fig.~\ref{compare}.
There is a clear peak due to the two-photon process
at small $| {\bm p}_{\mathrm T}^{*{\rm tot}} |$ values in Fig.~\ref{compare}~(a).
Good agreement is seen between the data and the signal MC 
for the $| {\bm p}_{\mathrm T}^{*{\rm tot}} |$ distribution.
Furthermore, the signal MC shows good agreement with the data sample
for the $|\cos\theta^{+-\gamma}_{\gamma}|$, $\cos\theta^{+-}_{-}$
and $\Delta \phi$ distributions. The simulation of angular distributions for the final state particles
can be performed well.
The overall signal detection efficiency in this analysis is estimated to be 7.36\%
using the signal MC.

\section{\boldmath Treatment of peaking background from ISR $\psi(2S)$ production}
\begin{figure}[htbp]
 \begin{minipage}{1.0\hsize}
  \begin{center}
 \includegraphics[width= 0.8\textwidth]{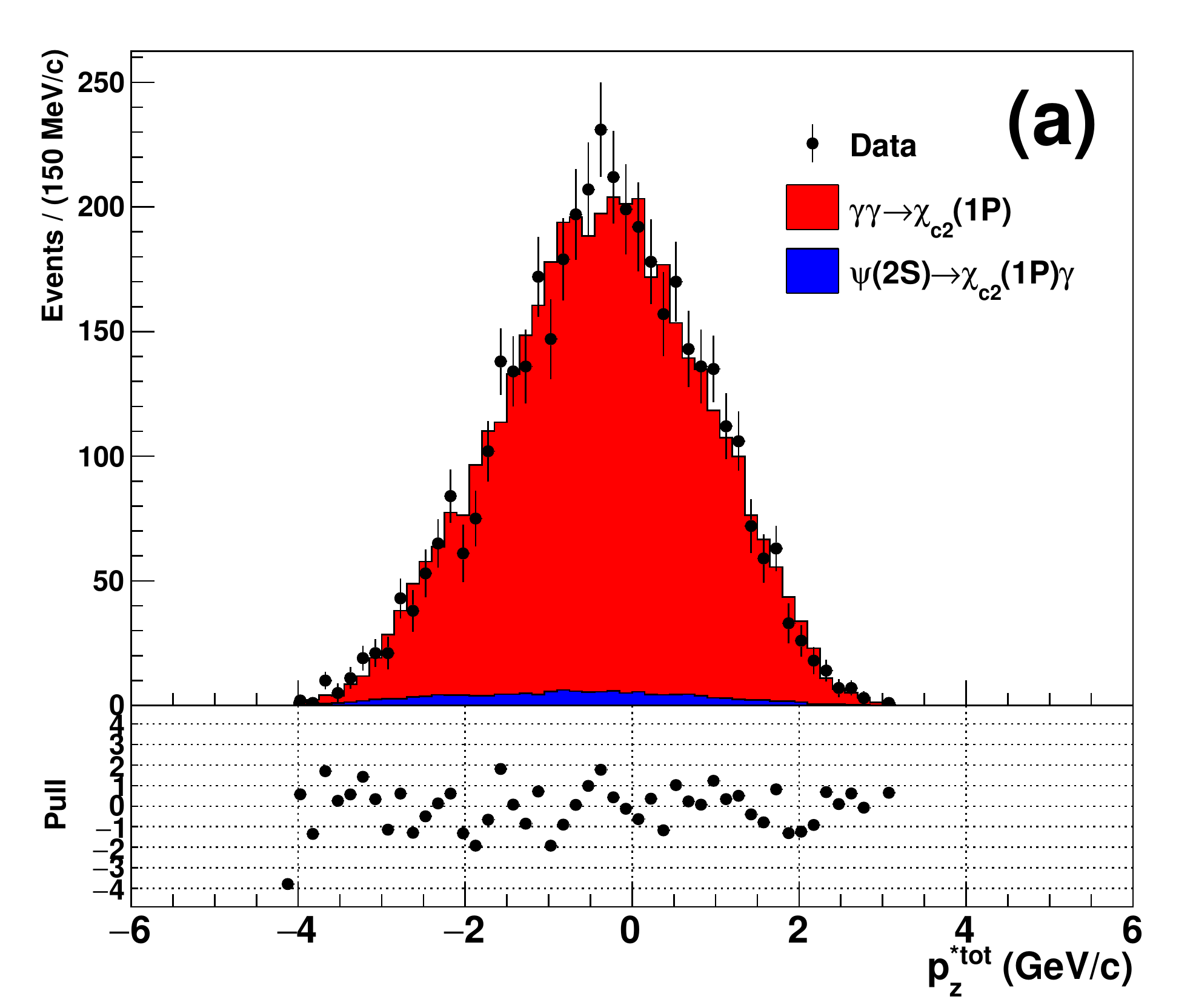}
  \end{center}
\end{minipage}
 \begin{minipage}{1.0\hsize}
  \begin{center}
  \includegraphics[width=0.8\textwidth]{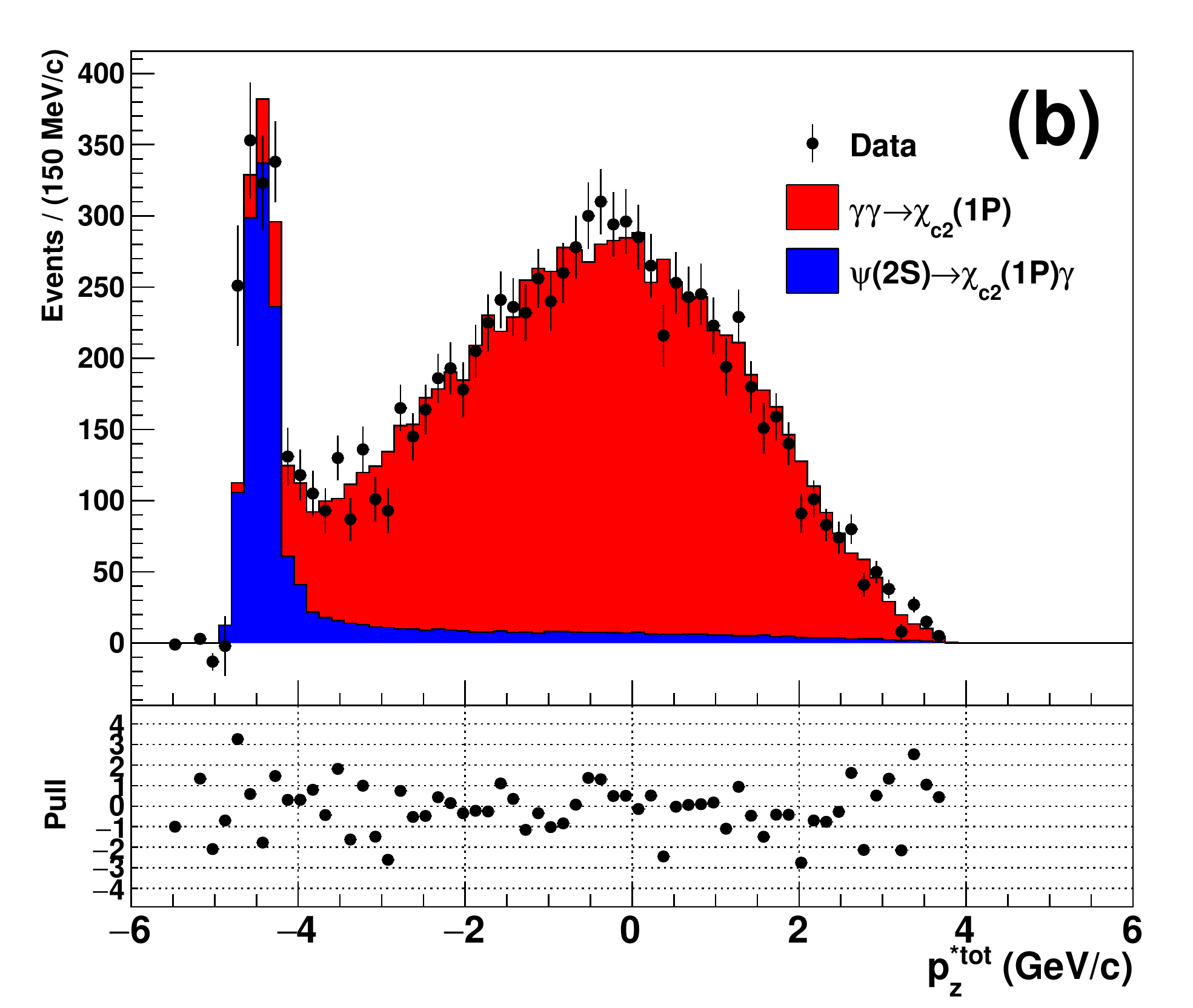}
  \end{center}
 \end{minipage}
  \caption{Distributions of $p_{z}^{*{\rm tot}}$
for the MC and 
the data sample, with background (estimated from $J/\psi$ sideband events)
subtracted (black points with error bars)
after (a) the standard event selection and (b) the loose event selection.
The peaking background component (blue histogram) in the MC is normalized 
based on the total expected number of peaking background events.
The signal component (red histogram) in the MC
is normalized based on the number of events in the peaking-background-subtracted data sample.}
  \label{pz}
 \end{figure}

A peaking background is anticipated from the process chain $e^{+} e^{-} \rightarrow \gamma_{ISR} \psi(2S),$ $\psi(2S) \rightarrow \chi_{c2}(1P) \gamma$, $\chi_{c2}(1P) \rightarrow J/\psi\gamma$, $J/\psi \rightarrow \ell^{+} \ell^{-}$. 
Events where the ISR photon and the photon from $\psi(2S)$ decay
are undetected have the same final state as the $\gamma \gamma  \rightarrow \chi_{c2}(1P)$ signal.
Therefore, we need to estimate the expected number of peaking background events.
The ISR $\psi(2S)$ production cross section has been precisely measured in Belle~\cite{psi_cross}.
To estimate the peaking background detection efficiency,
we prepare the peaking background MC
by using the PHOKHARA generator~\cite{phokhara}.
The requirements for angles of charged tracks and square of recoil mass
mostly remove the ISR events.
However, some double-ISR events where an ISR photon is emitted from each
beam, remain as their topology is similar to two-photon collision events.
The peaking background detection efficiency at the $\Upsilon(4S)$
is evaluated to be 0.55\% in this analysis.
The expected number of peaking background events and its uncertainty are estimated using
the ISR $\psi(2S)$ production cross section from the Belle study, the product of the relevant
branching fractions and the peaking background detection efficiency.
Taking the data sample with the different beam energies into account,
the total expected number of peaking background events
is estimated to be 170.9 $\pm$ 9.5 events:
the proportion of peaking background events in the $\chi_{c2}(1P)$ signal yield
is only 3.3\%.
The influence of the peaking background is not visible in any of the signal candidate distributions.
We finally evaluate the observed number of $\chi_{c2}(1P)$ signal events in two-photon processes by subtracting the total expected number of peaking background events
from the $\chi_{c2}(1P)$ signal yield.\\
\indent
To check the accuracy of the total expected number of peaking background events, 
we compare the peaking background MC and the data sample for 
the $p_{z}^{*{\rm tot}}$ distribution with the ``loose'' event selection,
where $p_{z}^{*{\rm tot}}$ is the sum of the $z$ component of momentum for the final state particles ($\ell^{+} \ell^{-} \gamma$) in the c.m.\ frame of the $e^{+}e^{-}$ beams.
In the loose event selection, only the requirement for the angles of the charged tracks ($-0.47 \leq \cos \theta \leq 0.82$) does not apply.
This requirement effectively rejects the peaking background 
and makes the characteristic peak due to the ISR process invisible on the $p_{z}^{*{\rm tot}}$ distribution.
Figure~\ref{pz} shows the $p_{z}^{*{\rm tot}}$ distribution in $\chi_{c2}(1P)$ signal candidates
for the MC, which consists of the signal MC and
the peaking background MC, and the background-subtracted data sample, where we have
applied a sideband subtraction similar to the original signal analysis,
changing the $J/\psi$ sideband region to 2.965~GeV/$c^{2}$ $<$ $M_{+-} <$ 3.000~GeV/$c^{2}$ and 3.150~GeV/$c^{2}$ $<$ $M_{+-} <$ 3.185~GeV/$c^{2}$,
for the $\Delta M$ region between 0.42~GeV/$c^{2}$ and 0.49~GeV/$c^{2}$
after the standard~(Fig.~\ref{pz}~(a)) and the loose event selection~(Fig.~\ref{pz}~(b)).
Unlike the original signal analysis, the distribution of the mass difference in the wide
sideband region is different from that in narrow sideband region for the peaking background process.
Therefore, we use the narrow sideband region.
The requirement for the angles of the charged tracks
removes 90.0\% of the ISR $\psi(2S) \rightarrow \chi_{c2}(1P) \gamma$ background
between Fig.~\ref{pz}~(a) and (b).
There is a clear peak due to the peaking background 
at about $-4.5$~GeV/$c$ on the $p_{z}^{*{\rm tot}}$ distribution in Fig.~\ref{pz}~(b),
and good agreement between the MC and the data sample in this
region, validating the estimate of the peaking background in $\Delta M$.

\section{Systematic uncertainties}
Table~\ref{sys1} shows the summary of systematic uncertainties
for the measurement of the two-photon decay width of $\chi_{c2}(1P)$.
We apply efficiency correction for lepton identification for the two tracks in the signal MC.
The lepton identification correction and its uncertainty
are estimated from a study of $e^{+}e^{-} \rightarrow e^{+}e^{-} \ell^{+} \ell^{-}$ 
events.
The $\chi_{c2}(1P)$ signal MC assumes pure helicity-2 production;
the associated systematic uncertainty is estimated by
varying the relevant parameter by its measured uncertainty, which
changes the $\ell^{+}\ell^{-}\gamma$ angular distribution (see Appendix \ref{appendixa}),
and noting the resulting change in the signal detection efficiency.
The uncertainty from the track finding efficiency 
is evaluated to be 0.3\% per track
using $ \tau \rightarrow \pi \pi^{0} \nu$, $\pi^{0} \rightarrow \gamma e^{+} e^{-}$ events. 
The uncertainty on the $J/\psi$ detection efficiency due to the definition of the $J/\psi$ signal region
is estimated by evaluating the difference in the detector resolution on the $M_{+-}$ distribution
between data and MC simulation. 
The systematic uncertainty for the photon detection efficiency is estimated
with radiative Bhabha events.
Since we require just one candidate for the signal photon from $\chi_{c2}(1P)$ decay in the event selection, signal events are rejected when we detect extra photons.
The uncertainty in the associated inefficiency is estimated
using the difference in the probability of detected extra photons
between data and MC simulation
using the $e^{+}e^{-} \rightarrow e^{+}e^{-} \mu^{+} \mu^{-}$ process with a
${\bm p}_{\mathrm T}^{*}$-balance requirement for $\mu^{+} \mu^{-}$.
The trigger efficiency estimated from the signal MC is 98.3\%.
We estimate the associated uncertainty by comparing the ratios of the different
sub-triggers between signal MC and experimental data.
We use the signal MC with only 10.58~GeV corresponding to $\Upsilon (4S)$ as $e^{+}e^{-}$ c.m.\ beam energy.
The effect of the different $e^{+}e^{-}$ beam energies is evaluated
based on the product of overall signal detection efficiency and luminosity function,
taking their luminosity-weighted average in the ratio to the $\Upsilon (4S)$ value.
The uncertainty due to neglecting the $\chi_{c2}(1P)$ total width
in the signal MC for the overall signal detection efficiency
is estimated using dedicated signal MC that takes the total width into account.
The validity and uncertainty of the $\Delta M$ fit method for $\chi_{c2}(1P)$ signal extraction
are estimated from a fitter test using toy MCs based on the shape of the signal MCs and
the experimental background component from the $J/\psi$ sideband events.
The estimated uncertainty includes
the fit bias of $\chi_{c2}(1P)$ signal yield and the uncertainty of the PDFs
estimated by fitter tests where the fixed shape parameters are varied.
The uncertainty on the luminosity function calculated by TREPS includes the effect of uncertainties
in the form factor, the radiative correction in the two-photon reaction,
and the difference in the approximation model for two-photon processes estimated 
from the comparison between TREPS and a full-diagram calculation \cite{aafh}
in the $e^{+}e^{-} \rightarrow e^{+}e^{-} \mu^{+} \mu^{-}$ process.
The systematic uncertainty on the integrated luminosity is estimated to be 1.4\%.
In addition to these sources,
we estimate the uncertainty on the photon energy resolution
which is applied to the signal photon in the signal MC.
The photon energy resolution correction and its uncertainty are estimated from a study
of $D^{*0} \rightarrow D^{0}\gamma$ events.
The effect on the overall signal detection efficiency is evaluated to be negligibly small.
The total systematic uncertainty is found to be 4.7\%, adding the various contributions in quadrature.

\begin{table}[htbp]
\begin{center}
\caption{Summary of the systematic uncertainties for $\Gamma_{\gamma\gamma}(\chi_{c2}(1P))$.}
 \begin{tabular}
 {@{\hspace{0.5cm}}l@{\hspace{0.5cm}}  @{\hspace{0.5cm}}c@{\hspace{0.5cm}}}
\hline \hline
Source &  Systematic uncertainty\\
\hline
Lepton ID efficiency correction & 0.8\%  \\
Angular distribution &  1.2\% \\
Tracking efficiency & 0.6\%  \\
$J/\psi$ detection efficiency & 2.4\%  \\ 
Photon detection efficiency & 2.0\%  \\
Inefficiency due to extra photons & 1.0\%  \\
Trigger efficiency & 0.9\%  \\
Different $e^{+}e^{-}$ beam energies & 0.1\%  \\
Neglecting  total width & 0.4\%  \\
Fit method & 0.6\%  \\
Luminosity function & 2.3\%  \\
Integrated luminosity & 1.4\%  \\ 
\hline
Total & 4.7\% \\
\hline \hline
 \end{tabular}
 \label{sys1}
\end{center}
\end{table}

\section{\boldmath Determination of two-photon decay width of $\chi_{c2}(1P)$}

\begin{figure}[htbp]
   \begin{center}
    \includegraphics[width=0.84\textwidth]{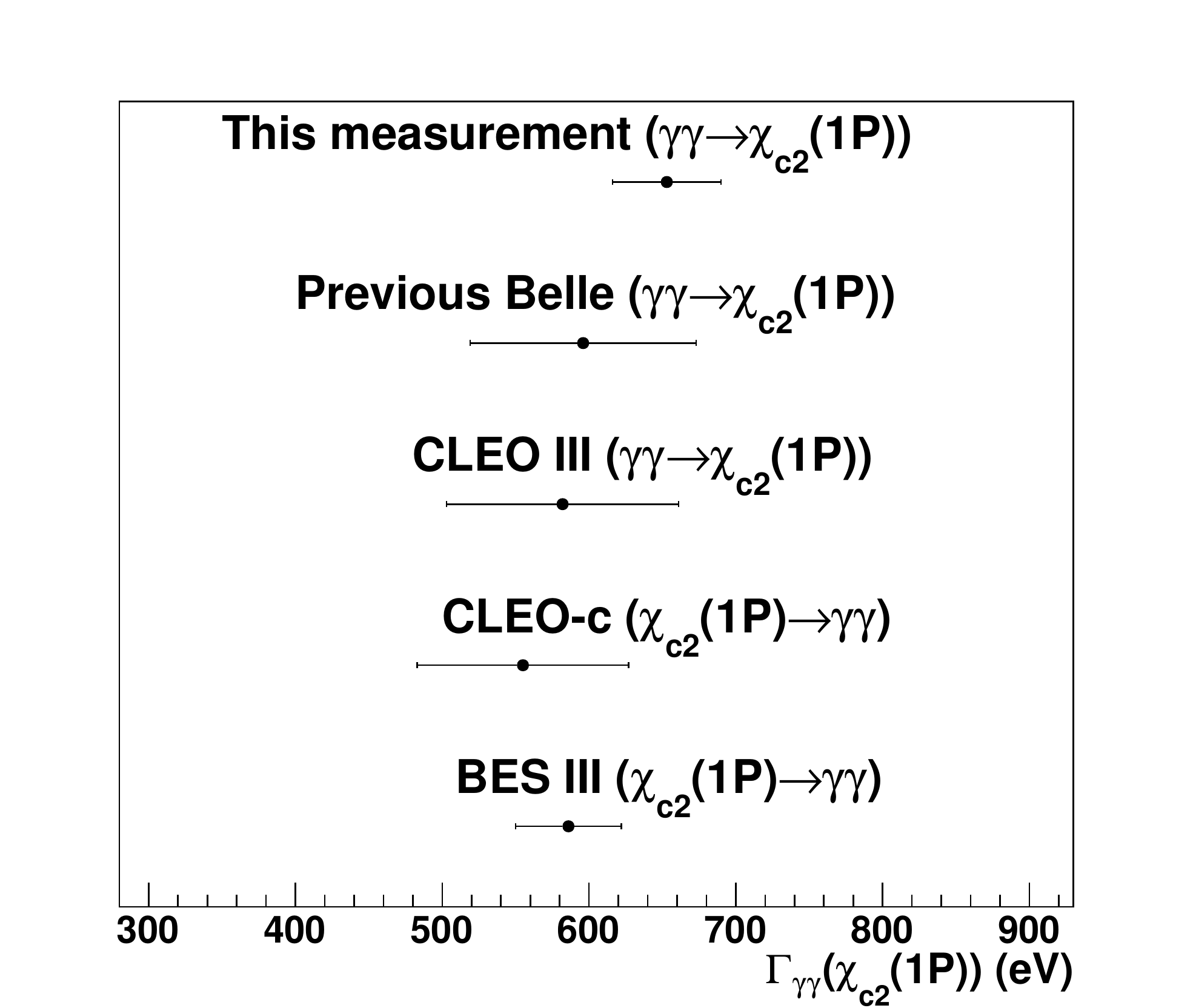}
    \caption{Comparison of experimental results for $\Gamma_{\gamma\gamma}(\chi_{c2}(1P))$,
where the error bars show the combined uncertainties from all sources added in quadrature.
This measurement supersedes the previous Belle result~\cite{previous_research}.}
    \label{mea_con2}
   \end{center}
  \end{figure}

Subtracting the total expected number of peaking background events
from the $\chi_{c2}(1P)$ signal yield,
the observed number of signal $\chi_{c2}(1P)$ events in two-photon processes is estimated to be 
4960.3 $\pm$ 97.9 events. 
From Eq.~\ref{eq:eq5}, the two-photon decay width of $\chi_{c2}(1P)$ is determined by
\begin{eqnarray}
\label{eq:eq6}
\Gamma_{\gamma \gamma}(\chi_{c2}(1P)) \mathcal{B}( \chi_{c2}(1P) \rightarrow J/\psi \, \gamma )\mathcal{B}( J/\psi \rightarrow \ell^+\ell^- ) = 
\frac{m_{\chi_{c2}(1P)}^2 N_{\rm sig}}{20 \pi^{2} (\int \mathcal{L} dt) \eta L_{\gamma \gamma}(m_{\chi_{c2}(1P)})}.
\end{eqnarray}
We substitute $m_{\chi_{c2}(1P)}$=3.556 GeV/$c^{2}$, ${\int \mathcal{L} dt}$=971~fb$^{-1}$,
$\eta$=7.36\% and $L_{\gamma \gamma}(m_{\chi_{c2}(1P)})=7.70\times10^{-4}$~GeV$^{-1}$, respectively.
From Eq.~\ref{eq:eq6}, the measured value is 
\begin{eqnarray}
\label{eq:eq7}
\Gamma_{\gamma \gamma}(\chi_{c2}(1P)) \mathcal{B}( \chi_{c2}(1P) \rightarrow J/\psi \, \gamma )\mathcal{B}( J/\psi \rightarrow \ell^+\ell^- ) = {\rm 14.8} \pm {\rm 0.3}({\rm stat.}) \pm {\rm 0.7}({\rm syst.}) \ {\rm eV}. \ \ \ \ \ 
\end{eqnarray}
This result corresponds to
\begin{eqnarray}
\label{eq:eq8}
\Gamma_{\gamma \gamma}(\chi_{c2}(1P)) = 653  \pm 13({\rm stat.}) \pm 31({\rm syst.}) \pm 17({\rm B.R.}) \ {\rm eV,}
\end{eqnarray}
where the third uncertainty is from $\mathcal{B}( \chi_{c2}(1P) \rightarrow J/\psi \, \gamma )$ = (19.0 $\pm$ 0.5)\% and $\mathcal{B}( J/\psi \rightarrow \ell^+\ell^- )$ = (11.93 $\pm$ 0.05)\%~\cite{pdg2020}.
Table~\ref{mea_con} and Fig.~\ref{mea_con2} show the summary and comparison of experimental results for $\Gamma_{\gamma\gamma}(\chi_{c2}(1P))$, respectively.
This measurement is the most precise measurement of $\Gamma_{\gamma\gamma}(\chi_{c2}(1P))$ 
in two-photon processes and consistent with the previous Belle result~\cite{previous_research} and 
the other experimental results~\cite{CLEOc,bes3,CLEO3}.
Precision of the present measurement is almost the same as
that of the most precise measurement from the $\chi_{c2}(1P)$ decay of \mbox{BES~I\hspace{-.1em}I\hspace{-.1em}I}~\cite{bes3}.

\begin{table*}[htb]
\begin{threeparttable}[h]
\caption{Summary of experimental results for $\Gamma_{\gamma\gamma}(\chi_{c2}(1P))$, where
$\mathcal{B}_{1} \equiv \mathcal{B}( \chi_{c2}(1P) \rightarrow J/\psi \, \gamma )$, $\mathcal{B}_{2} \equiv \mathcal{B}( J/\psi \rightarrow \ell^+\ell^- )$, $\mathcal{B}_{3} \equiv \mathcal{B}( \psi(2S) \rightarrow \chi_{c2}(1P) \gamma )$, $\mathcal{B}_{4} \equiv \mathcal{B}(\chi_{c2}(1P) \rightarrow \gamma \gamma)$.}
\label{mea_con}
\begin{tabular}{lcc}
\hline \hline
Experiment [Ref.] & Measured value & $\Gamma_{\gamma \gamma}(\chi_{c2}(1P))$ (eV)\\
\hline
This measurement \ \ &  \ \ $\Gamma_{\gamma \gamma}(\chi_{c2}(1P)) \times \mathcal{B}_{1} \times \mathcal{B}_{2} = 14.8 \pm 0.3 \pm 0.7$ eV \ \  &  \ $ 653 \pm13 \pm 31 \pm 17\tnote{a} \ $ \\
Previous Belle \cite{previous_research} &  $\Gamma_{\gamma \gamma}(\chi_{c2}(1P)) \times \mathcal{B}_{1} \times \mathcal{B}_{2} = 13.5 \pm 1.3 \pm 1.1$ eV  &  $ 596 \pm 58 \pm 48 \pm 16\tnote{a,b} $ \\
\mbox{CLEO~I\hspace{-.1em}I\hspace{-.1em}I} \cite{CLEO3} &  $\Gamma_{\gamma \gamma}(\chi_{c2}(1P)) \times \mathcal{B}_{1} \times \mathcal{B}_{2} = 13.2 \pm 1.4 \pm 1.1$ eV  &  $ 582 \pm 59 \pm 50 \pm 15\tnote{a,b} $ \\
CLEO-c \cite{CLEOc} & $ \mathcal{B}_{3} \times \mathcal{B}_{4} \times 10^{5} = 2.68  \pm 0.28 \pm 0.15 $ & $ 555 \pm 58 \pm 32 \pm 28\tnote{c,d} $ \\
\mbox{BES~I\hspace{-.1em}I\hspace{-.1em}I} \cite{bes3} & $ \mathcal{B}_{3} \times \mathcal{B}_{4} \times 10^{5} = 2.83  \pm 0.08 \pm 0.06 $ & $ 586 \pm 16 \pm 13 \pm 29\tnote{c,d} $ \\
\hline \hline
\end{tabular}
\begin{tablenotes}
\item[a] Third uncertainty is associated with the uncertainties of $\mathcal{B}( \chi_{c2}(1P) \rightarrow J/\psi \, \gamma )$ and $\mathcal{B}( J/\psi \rightarrow \ell^+\ell^- )$.
\item[b] The results is recalculated by using $\mathcal{B}( \chi_{c2}(1P) \rightarrow J/\psi \gamma ) = (19.0 \pm 0.5) \%$ and $\mathcal{B}(J/\psi \rightarrow \ell^{+} \ell^{-}) = (11.93 \pm 0.05) \%$ from PDG \cite{pdg2020}.
\item[c] Third uncertainty is associated with the uncertainties of $\mathcal{B}(\psi(2S) \rightarrow \chi_{c2}(1P) \gamma)$ and the total width of $\chi_{c2}(1P)$.
\item[d] The results is recalculated by using $\mathcal{B}( \psi(2S) \rightarrow \chi_{c2}(1P) \gamma ) = (9.52 \pm 0.20) \%$ and $\Gamma_{\chi_{c2}(1P)} = 1.97 \pm 0.09$ MeV from PDG \cite{pdg2020}.
\end{tablenotes}
\end{threeparttable}
\end{table*}

\section{Conclusion}

We measure the two-photon decay width of $\chi_{c2}(1P)$
in the analysis of $ \gamma \gamma \rightarrow \chi_{c2}(1P) \rightarrow J/\psi\gamma$,
$J/\psi \rightarrow \ell^{+} \ell^{-} $ $(\ell = e \ {\rm or} \ \mu)$
using the 971~fb$^{-1}$ data sample at Belle.
The measured value is 
$\Gamma_{\gamma \gamma}(\chi_{c2}(1P)) \mathcal{B}( \chi_{c2}(1P) \rightarrow J/\psi \, \gamma )\mathcal{B}( J/\psi \rightarrow \ell^+\ell^- ) = {\rm 14.8} \pm {\rm 0.3}({\rm stat.}) \pm {\rm 0.7}({\rm syst.}) \ {\rm eV}$.
This result corresponds to $ \Gamma_{\gamma \gamma}(\chi_{c2}(1P)) $
 = 653  $\pm$ 13(stat.) $\pm$ 31(syst.) $\pm$ 17(B.R.)~eV,
where the third uncertainty is from $\mathcal{B}( \chi_{c2}(1P) \rightarrow J/\psi \, \gamma )$ and $\mathcal{B}( J/\psi \rightarrow \ell^+\ell^- )$.
The result in this paper is the most precise measurement of $\Gamma_{\gamma\gamma}(\chi_{c2}(1P))$
in two-photon processes and has a compatible precision with that from
the $\chi_{c2}(1P)$ decay of \mbox{BES~I\hspace{-.1em}I\hspace{-.1em}I}~\cite{bes3}.
This measurement supersedes the previous Belle result~\cite{previous_research}.

\appendix

\section{Angular distribution}
\label{appendixa}

In the decay process $\chi_{c2}(1P) \rightarrow J/\psi \gamma$,
E1, M2 and E3 transitions are allowed.
Furthermore, the helicity of $\chi_{c2}(1P)$ with respect to the $\gamma$$\gamma$ axis
can have $\uplambda=0$ or 2 in the production process.
Taking these conditions in $\gamma \gamma  \rightarrow \chi_{c2}(1P)$, $\chi_{c2}(1P) \rightarrow J/\psi\gamma$, $J/\psi \rightarrow \ell^{+} \ell^{-}$ into account, 
the normalized angular distribution of the final state, which is written as 
$\hat{W}(\theta,\theta^{*},\phi^{*})$, is given by Eq.~\ref{eq:eq9}.
In this equation, $\theta$ is the polar angle of the photon from $\chi_{c2}(1P)$ decay with respect to the $\gamma \gamma$ axis in the 
$\ell^{+} \ell^{-} \gamma$ c.m.\ frame; $\theta ^{*} $ and $\phi^{*}$ are the 
polar and  azimuthal angles of  
$ \ell^{-} $, respectively, in the $x$-$z$ plane of the $J/\psi$ and $ \ell^{+} \ell^{-} $ 
c.m.\ frame, where the $z$-axis is along the direction of the $J/\psi$ and
the $x$-axis is in the $J/\psi\gamma$ scattering plane;
$A_{k}$ shows the helicity amplitudes of $\chi_{c2}(1P)$ with respect to the $\gamma$$J/\psi$ axis, where $k$ is the absolute value of helicity; $a_{1}$, $a_{2}$ and $a_{3}$ are the fractional multipole amplitudes of E1, M2 and E3 transitions, respectively.
The relationship between $A_{k}$ and $a_{i}$ is also shown;
$A_{k}$ and $a_{i}$ are normalized;
the sign of $a_{1}$ is conventionally defined to be positive; and
$R$ is the ratio of $\uplambda=2$ to the whole ($\uplambda=0$ or $2$).\\
\indent
In the signal MC,
we assume $a_{2}=-0.11 \pm 0.01$ and $a_{3}=0.00 \pm 0.01$~\cite{pdg2020}
for the $\chi_{c2}(1P) \rightarrow J/\psi \gamma$.
For the state of helicity of $\chi_{c2}(1P)$ with
respect to the $\gamma$$\gamma$ axis,
we assume a pure $\uplambda=2$ state, based on the measurement $R=1.000 \pm 0.018$
by \mbox{BES~I\hspace{-.1em}I\hspace{-.1em}I} experiment\footnotemark[4]\footnotetext[4]{We take the linear sum of the statistical and systematic uncertainties, conservatively, for our uncertainty evaluation.}~\cite{bes3}.
\begin{eqnarray}
\label{eq:eq9}
\frac{64\pi^{2}}{15}\hat{W}(\theta,\theta^{*},\phi^{*})=K_{1}+K_{2} \cos^{2}\theta +  K_{3} \cos^{4}\theta 
 + (K_{4}+K_{5} \cos^{2}\theta +  K_{6} \cos^{4}\theta)\cos^{2}\theta^{*} \ \ \ \ \ \   \notag \\
 +  (K_{7}+K_{8} \cos^{2}\theta +  K_{9} \cos^{4}\theta) \sin^{2}\theta^{*} \cos 2 \phi^{*} 
 +  (K_{10}+K_{11} \cos^{2}\theta) \sin 2 \theta \sin 2 \theta^{*} \cos \phi^{*} \ \ \ \ \ \   
\end{eqnarray}

\begin{eqnarray*}
\begin{cases}
16K_{1}=4A^{2}_{0}+6A^{2}_{2}+R (2A^{2}_{0}+8A^{2}_{1}-5A^{2}_{2})\\
8K_{2}=3(-4A^{2}_{0}+8A^{2}_{1}-2A^{2}_{2}+R(2A^{2}_{0}-8A^{2}_{1}+3A^{2}_{2}))\\
16K_{3}=(6A^{2}_{0}-8A^{2}_{1}+A^{2}_{2})(6-5R)\\
16K_{4}=4A^{2}_{0}+6A^{2}_{2}+R(2A^{2}_{0}-8A^{2}_{1}-5A^{2}_{2})\\
8K_{5}=3(-4A^{2}_{0}-8A^{2}_{1}-2A^{2}_{2}+R(2A^{2}_{0}+8A^{2}_{1}+3A^{2}_{2}))\\
16K_{6}=(6A^{2}_{0}+8A^{2}_{1}+A^{2}_{2})(6-5R)\\
8K_{7}=\sqrt{6}(3R-2)A_{0}A_{2}\\
K_{8}=\sqrt{6}(1-R)A_{0}A_{2}\\
8K_{9}=\sqrt{6}(5R-6)A_{0}A_{2}\\
8K_{10}=\sqrt{3}(2A_{0}A_{1}+\sqrt{6}A_{1}A_{2}-R(A_{0}A_{1}+3\sqrt{3/2}A_{1}A_{2}))\\
8\sqrt{3}K_{11}=(5R-6)(3A_{0}A_{1}+\sqrt{3/2}A_{1}A_{2})\\
\end{cases}
\end{eqnarray*}

\begin{flalign*}
\sum_{k=0}^{2}A^{2}_{k} = \sum_{i=1}^{3}a^{2}_{k} = 1
\end{flalign*}
\begin{flalign*}
\begin{cases}
A_{0}=\sqrt{\frac{1}{10}}a_{1} + \sqrt{\frac{1}{2}}a_{2} + \sqrt{\frac{6}{15}}a_{3}\\
A_{1}=\sqrt{\frac{3}{10}}a_{1} + \sqrt{\frac{1}{6}}a_{2} - \sqrt{\frac{8}{15}}a_{3}\\
A_{2}=\sqrt{\frac{6}{10}}a_{1} - \sqrt{\frac{1}{3}}a_{2} + \sqrt{\frac{1}{15}}a_{3}\\
\end{cases}
\end{flalign*}

\section*{Acknowledgements}

This work, based on data collected using the Belle detector, which was
operated until June 2010, was supported by 
the Ministry of Education, Culture, Sports, Science, and
Technology (MEXT) of Japan, the Japan Society for the 
Promotion of Science (JSPS), and the Tau-Lepton Physics 
Research Center of Nagoya University; 
the Australian Research Council including grants
DP180102629, 
DP170102389, 
DP170102204, 
DE220100462, 
DP150103061, 
FT130100303; 
Austrian Federal Ministry of Education, Science and Research (FWF) and
FWF Austrian Science Fund No.~P~31361-N36;
the National Natural Science Foundation of China under Contracts
No.~11675166,  
No.~11705209;  
No.~11975076;  
No.~12135005;  
No.~12175041;  
No.~12161141008; 
Key Research Program of Frontier Sciences, Chinese Academy of Sciences (CAS), Grant No.~QYZDJ-SSW-SLH011; 
Project ZR2022JQ02 supported by Shandong Provincial Natural Science Foundation;
the Ministry of Education, Youth and Sports of the Czech
Republic under Contract No.~LTT17020;
the Czech Science Foundation Grant No. 22-18469S;
Horizon 2020 ERC Advanced Grant No.~884719 and ERC Starting Grant No.~947006 ``InterLeptons'' (European Union);
the Carl Zeiss Foundation, the Deutsche Forschungsgemeinschaft, the
Excellence Cluster Universe, and the VolkswagenStiftung;
the Department of Atomic Energy (Project Identification No. RTI 4002) and the Department of Science and Technology of India; 
the Istituto Nazionale di Fisica Nucleare of Italy; 
National Research Foundation (NRF) of Korea Grant
Nos.~2016R1\-D1A1B\-02012900, 2018R1\-A2B\-3003643,
2018R1\-A6A1A\-06024970, RS\-2022\-00197659,
2019R1\-I1A3A\-01058933, 2021R1\-A6A1A\-03043957,
2021R1\-F1A\-1060423, 2021R1\-F1A\-1064008, 2022R1\-A2C\-1003993;
Radiation Science Research Institute, Foreign Large-size Research Facility Application Supporting project, the Global Science Experimental Data Hub Center of the Korea Institute of Science and Technology Information and KREONET/GLORIAD;
the Polish Ministry of Science and Higher Education and 
the National Science Center;
the Ministry of Science and Higher Education of the Russian Federation, Agreement 14.W03.31.0026, 
and the HSE University Basic Research Program, Moscow; 
University of Tabuk research grants
S-1440-0321, S-0256-1438, and S-0280-1439 (Saudi Arabia);
the Slovenian Research Agency Grant Nos. J1-9124 and P1-0135;
Ikerbasque, Basque Foundation for Science, Spain;
the Swiss National Science Foundation; 
the Ministry of Education and the Ministry of Science and Technology of Taiwan;
and the United States Department of Energy and the National Science Foundation.
These acknowledgements are not to be interpreted as an endorsement of any
statement made by any of our institutes, funding agencies, governments, or
their representatives.
We thank the KEKB group for the excellent operation of the
accelerator; the KEK cryogenics group for the efficient
operation of the solenoid; and the KEK computer group and the Pacific Northwest National
Laboratory (PNNL) Environmental Molecular Sciences Laboratory (EMSL)
computing group for strong computing support; and the National
Institute of Informatics, and Science Information NETwork 6 (SINET6) for
valuable network support.


\begin{thebibliography}{99}


\bibitem{nonrelative_2}
R. Barbier, R. Gatto, and R. Kögerler, Phys. Lett. B {\bf 60}, 183 (1976)

\bibitem
{relativistic_corrections_1}
C.R.M\"{u}nz, Nucl. Phys. A {\bf 609}, 364 (1996).

\bibitem{relativistic_quark_model_2}
S. Godfrey and N. Isgur, Phys. Rev. D {\bf 32}, 189 (1985).

\bibitem{potential}
S. N. Gupta, J. M. Johnson, and W. W. Repko, Phys. Rev. D {\bf 54}, 2075 (1996).

\bibitem{relativistic_quark_model_1}
D. Ebert, R. N. Faustov, and V. O. Galkin, Mod. Phys. Lett. A {\bf 18}, 601 (2003).

\bibitem{rigorous_QCD_2}
G. T. Bodwin, E. Braaten, and G. P. Lepage, Phys. Rev. D {\bf 46}, R1914 (1992).

 \bibitem{rigorous_QCD_1}
H. W. Huang and K. T. Chao, Phys. Rev. D {\bf 54}, 6850 (1996); {\bf 56}, 1821(E) (1997).

\bibitem{nonrelativistic_QCD_factorization_framework}
G. A. Schuler, F. A. Berends, and R. van Gulik, Nucl. Phys. B {\bf 523}, 423 (1998).

\bibitem{two-body_Dirac_equations_of_constraint_dynamics}
H. W. Crater, C. Y. Wong and P. Van Alstine, Phys. Rev. D {\bf 74}, 054028 (2006).

\bibitem{effective_Lagrangian}
J. P. Lansberg and T. N. Pham, Phys. Rev. D {\bf 79}, 094016 (2009).

\bibitem{light_front}
C. W. Hwang and R. S. Guo, Phys. Rev. D {\bf 82}, 034021 (2010).

\bibitem{CLEOc}
K. M. Ecklund {\it et al.} (CLEO Collaboration), Phys. Rev. D {\bf 78}, 091501(R) (2008).

\bibitem{bes3}
M. Ablikim {\it et al.} (\mbox{BES~I\hspace{-.1em}I\hspace{-.1em}I} Collaboration), Phys. Rev. D {\bf 96}, 092007 (2017).

\bibitem{previous_research}
K. Abe {\it et al.} (Belle Collaboration), Phys. Lett. B {\bf 540}, 33 (2002).

\bibitem{CLEO3}
S. Dobbs {\it et al.} (CLEO Collaboration), Phys. Rev. D {\bf 73}, 071101(R) (2006).

\bibitem{pdg2020}
P.A. Zyla {\it et al.} (Particle Data Group), Prog. Theor. Exp. Phys. {\bf 2020}, 083C01 (2020).

\bibitem{epa}
C. Berger and W. Wagner, Phys. Rept. {\bf 146}, 1 (1987).

\bibitem{belledetector1}
A. Abashian {\it et al.} (Belle Collaboration), Nucl. Instr. and Methods Phys. Res. Sect. A {\bf 479}, 117 (2002).

\bibitem{belledetector2}
J. Brodzicka {\it et al.}, Prog. Theor. Exp. Phys. {\bf 2012}, 04D001 (2012).

\bibitem{KEKB1}
S. Kurokawa and E. Kikutani, Nucl. Instr. and Methods Phys. Res. Sect. A {\bf 499}, 1 (2003), and other papers included in this volume.

\bibitem{KEKB2}
T. Abe {\it et al.}, Prog. Theor. Exp. Phys. {\bf 2013}, 03A001 (2013) and following articles up to 03A011.

\bibitem{treps}
S.Uehara, KEK Report 96-11, arXiv:1310.0157[hep-ph] (1996).

\bibitem{photos}
E. Barberio, Z. W\c{a}s, Comput. Phys. Commun. {\bf 79} (1994) 291.

\bibitem{geant3}
R. Brun {\it et al.}, GEANT3.21, CERN Report No. DD/EE/84-1 (1987).

\bibitem{teramoto}
Y. Teramoto {\it et al.} (Belle Collaboration), Phys. Rev. Lett. {\bf 126}, 122001 (2021).

\bibitem{chi_b}
V. Bhardwaj {\it et al.} (Belle Collaboration), Phys. Rev. D {\bf 93}, 052016 (2016).

\bibitem{psi_cross}
Z. Q. Liu {\it et al.} (Belle Collaboration), Phys. Rev. Lett. {\bf 110}, 252002 (2013).

\bibitem{phokhara}
G. Rodrigo {\it et al.}, Eur. Phy. J. C {\bf 24} (2002) 71-82.

\bibitem{aafh}
F.A. Berends, P.H. Daverveldt and R. Kleiss, Comput. Phys. Commun. {\bf 40}, 285 (1986).






\end{thebibliography}
\end{document}